\documentclass[fleqn,usenatbib,twocolumn]{mn2e}

\usepackage[T1]{fontenc}
\usepackage{ae,aecompl}
\usepackage{graphicx}
\usepackage{dcolumn}
\usepackage{amsmath}
\usepackage{amssymb}
\usepackage{color}
\usepackage{enumerate}

\newcommand{\aap}{{Astron.~Astrophys.}}
\newcommand{\apjl}{{Astrophys.~J.~Lett.}}
\newcommand{\apjs}{{Astrophys.~J.~Supp.}}

\newcommand{\be}{\begin{equation}}
\newcommand{\ee}{\end{equation}}
\newcommand{\bea}{\begin{eqnarray}}
\newcommand{\eea}{\end{eqnarray}}

\newcommand{\GeV}{{\rm ~GeV}}
\newcommand{\TeV}{{\rm ~TeV}}

\def\la{\vcenter{\hbox{$<$}\offinterlineskip\hbox{$\sim$}}}
\def\ga{\vcenter{\hbox{$>$}\offinterlineskip\hbox{$\sim$}}}

\title{A Latitude-Dependent Analysis of the Leptonic Hypothesis for the \emph{Fermi} Bubbles}

\author[S. A. Narayanan \& T. R. Slatyer]{
Sruthi A. Narayanan\thanks{E-mail: sruthian@mit.edu},
T. R. Slatyer\thanks{E-mail: tslatyer@mit.edu}
\\
Center for Theoretical Physics, Massachusetts Institute of Technology, Cambridge, MA 02139, USA\\
}

\date{Accepted XXX. Received YYY; in original form ZZZ}

\pubyear{2016}

\begin{document}
\maketitle

\begin{abstract} The \emph{Fermi} Bubbles are giant Galactic structures observed in both gamma-rays and microwaves. Recent studies have found support for the hypothesis that the gamma-ray and microwave emission can both be understood as arising from a hard cosmic-ray electron population within the volume of the Bubbles, via inverse Compton scattering and synchrotron radiation respectively. The relative rates of these processes are set by the relative energy density of the interstellar radiation field and the magnetic field within the Bubbles; consequently, under the hypothesis of a common origin, the combination of the gamma-ray and microwave measurements can be used to estimate the magnetic field within the Bubbles. We revisit the consistency of this hypothesis on a latitude-by-latitude basis, using data from \emph{Fermi}, WMAP and \emph{Planck}; estimate the variation of the electron spectrum within the Bubbles; and infer bounds on the magnetic field within the Bubbles as a function of distance from the Galactic plane. We find that while the microwave and gamma-ray spectra are generally consistent with the leptonic hypothesis for few-microGauss magnetic fields, there appears to be a preference for spectral hardening in the microwaves at mid-latitudes (especially in the $|b| \sim 25-35^\circ$ range) that is not mirrored in the gamma rays. This result may hint at a non-leptonic contribution to the gamma-ray spectra; however, the discrepancy can be reconciled in purely leptonic models if the cutoff energy for the electrons is lower in this latitude range and the spectrum below the cutoff is harder. \end{abstract}




\section{Introduction}

Observations of the gamma-ray sky by the \emph{Fermi} Gamma-Ray Space Telescope have revealed a spectrally hard bilobular structure centred at the Galactic Centre (GC), and extending to Galactic latitudes of $\sim \pm 50^\circ$ \citep{Su:2010qj}. The initial study found that the gamma-ray spectrum of these `\emph{Fermi} Bubbles' was approximately spatially invariant in both shape and amplitude; subsequent analyses \citep{2012ApJ...753...61S, Hooper:2013rwa, Yang:2014pia, apj2014_14077905} of these \emph{Fermi} Bubbles have confirmed this general statement, while finding some evidence for local spectral variation. In particular, \cite{Yang:2014pia} found evidence for hardening of the low-energy spectrum at the southern end of the Bubbles (see also \cite{Selig:2014qqa}).

The search for the Bubbles was originally motivated by the presence of the `\emph{WMAP} haze', an excess of relatively hard microwave emission from the region surrounding the GC, discovered using \emph{WMAP} data in 2003 \citep{Finkbeiner:2003im}. Under the `synchrotron hypothesis', the \emph{WMAP} haze could be explained by a new population of hard-spectrum electrons in this region. Such a population should give rise to a corresponding signal in gamma rays, via inverse Compton scattering (ICS) of the ambient interstellar radiation field (ISRF) by the same hard electrons. Various analyses have demonstrated that for few-microGauss magnetic fields, it is possible to simultaneously explain the overall spectrum of the \emph{Fermi} Bubbles and the microwave haze with the same electron population \citep{Dobler:2009xz, Su:2010qj, apj2014_14077905}. Studies using data from  \emph{Planck}  have both confirmed the existence of a microwave haze and found striking spatial overlap with the gamma-ray bubbles \citep{2013A&A...554A.139P, Dobler:2012ef}. 

Possible counterparts to the Bubbles have also been claimed at other frequencies -- notably in polarized radio \citep{Carretti:2013sc} and X-ray data \citep{Su:2010qj, Tahara:2015mia, Kataoka:2015dla}. Ultraviolet absorption-line spectra have provided evidence for an expanding biconical outflow from the GC, produced over the past $\sim 2.5-4.0$ Myr \citep{2015ApJ...799L...7F}.\footnote{However, this might reflect an internal flow within an older structure, rather than the original formation of the Bubbles; we thank Roland Crocker for this point.} However, in this work we will focus on the comparison between the microwave and gamma-ray data.

The nature and origin of the \emph{Fermi} Bubbles remains an open question and subject of active research. Star formation or starburst activity, or a jet from the black hole, are the main candidates for the energy sources (although it has been argued that modified diffusion within the Bubbles can explain the gamma-ray signal without requiring a new source of particle production \citep{Thoudam:2013eaa}). Starbursts may steadily inject high-energy protons into the region of the Bubbles \citep{2011PhRvL.106j1102C}, or launch winds \citep{2011MNRAS.413..763C} whose termination shock accelerates particles to high energies \citep{Lacki:2013zsa}; similar winds could be fueled by a hot accretion phase of the supermassive black hole at the GC \citep{Mou:2014pea, Mou:2015wxa}. The multiwavelength signatures of such winds have been recently studied by \cite{Sarkar:2015xta}. Bubbles arising from jets have been simulated by several authors \citep{2012ApJ...756..181G, 2012ApJ...756..182G,2012ApJ...761..185Y}; more recently, detailed simulations of the leptons in Active Galactic Nuclei (AGN) jets and the resulting inverse Compton and synchrotron signals have been carried out \citep{2013MNRAS.436.2734Y}. Stochastic Fermi acceleration in turbulent magnetic fields, inside the Bubbles, may yield the requisite population of high-energy electrons \citep{Mertsch:2011es, 2014ApJ...790...23C, Cheng:2015zda}. The possible formation of a large-scale magnetic structure associated with the Bubbles, via an explosive event, has been studied by \cite{Barkov:2013gda}; these authors suggest that such an effect might explain the high-latitude hardening claimed by \cite{Yang:2014pia}.

The advent of improved simulations and theoretical models for the Bubbles motivates a reconsideration of the raw data, and what it can teach us. In particular, if the microwave and gamma-ray signals indeed originate from the same electron population (as suggested by the degree of spatial coincidence between them), then it is possible to simultaneously (1) estimate the consistency of the electron spectrum required to fit the two signals and (2) estimate the magnetic field within the Bubbles, on a latitude-by-latitude basis. These questions have already been studied by \cite{Hooper:2013rwa}, but that paper focused primarily on the discovery of a new gamma-ray emission component at low latitude. In this work we present a more detailed and careful analysis, finding that the spectra are consistent on a latitude-by-latitude basis, albeit the microwave data suggest a spectral hardening at mid-high latitudes that in turn requires a high-energy cutoff in the electron spectrum in order to fit the gamma-ray data.

In contrast to the originally proposed leptonic scenario, where the gamma-ray Bubbles and microwave haze originate from the same electron population via ICS and synchrotron respectively, many of the models described above posit a hadronic origin for the gamma-ray emission (see e.g. \cite{2011PhRvL.106j1102C, Fujita:2013jda}). In such scenarios, high-energy protons interact with the ambient gas and produce neutral pions, which decay to gamma rays. This process also yields charged pions, which decay to electrons and/or positrons. However, the synchrotron radiation from these secondary electrons and positrons cannot generally yield the microwave haze, for the same parameters that match the gamma-ray emission \citep{Fujita:2014oda, apj2014_14077905, 2015ApJ...799..112C, Sarkar:2015xta}. Consequently, models that are not purely leptonic are generally mixed, including both a hadronic contribution and some source of primary electrons (e.g. \cite{Crocker:2014fla}). 

In this work we will generally assume a purely leptonic scenario; a partially hadronic scenario would  relax the constraints on the electron spectrum (as it would no longer be necessary to match the gamma-ray spectrum), while preferring higher magnetic fields (so that the reduced primary leptonic component could still generate the microwave haze). Thus our estimates for magnetic field strength should be taken as a lower bound, and adjusted appropriately in the presence of a substantial hadronic contribution.

In Section \ref{sec:methods} we begin by describing the datasets we use, reviewing the necessary results for ICS and synchrotron radiation from high-energy electrons, and outlining the methods we employ. In Section \ref{sec:overallcomp} we reproduce results from the existing literature as a cross-check, and discuss the analysis of the Bubbles as a whole, before moving on to study the spectra preferred independently by the microwave and gamma-ray data on a latitude-dependent basis in Section \ref{sec:latdep}. We present the combined latitude-dependent analysis, and the implications for the latitude-dependent magnetic field, in Section \ref{sec:simultaneous}. Finally we present our conclusions in Section \ref{sec:conclusion}.  In Appendix~\ref{app:hardcutoff} we discuss results using a hard cutoff model for the electron spectrum, to test sensitivity to the modeling of the electron spectrum at high energies; in Appendix \ref{app:innergalaxy} we present a focused discussion of the low-latitude region where the background modeling becomes particularly uncertain; we test the effects of some other possible systematics in Appendix \ref{app:systematics}, and explore the effects of varying the ISRF model in Appendix \ref{app:isrf}.
 
\section{Data and Methods}
\label{sec:methods}

\subsection{Gamma-ray data from the \emph{Fermi} Gamma-Ray Space Telescope}

For the gamma-ray waveband, we employ data from the \emph{Fermi} Gamma-Ray Space Telescope (hereafter \emph{Fermi}). \emph{Fermi} is capable of detecting gamma rays with energies between $\sim20$ MeV and several TeV; for this analysis we examine only data from $0.1-500$ GeV, and further restrict the energy range for some analyses.

\subsubsection{Fermi Data Set 1: \emph{Fermi} Collaboration Analysis of the Bubbles}

The \emph{Fermi} Collaboration has presented a careful study of the Bubbles \citep{apj2014_14077905}, including an estimate of the latitude-dependent emission in three broad latitude ranges ($10-20^\circ$, $20-40^\circ$ and $40-60^\circ$), in the northern and southern sky separately. These results cover the full energy range from 0.1-500 GeV, with 25 logarithmically spaced energy bins.

The principal challenge in extracting the spectrum of the Bubbles is modeling of the diffuse gamma-ray background, arising from interactions between cosmic ray nuclei and interstellar gas, and ICS and bremsstrahlung emission from cosmic ray electrons. The distribution of gamma rays from these sources can be modelled using gas maps and numerical models for the distribution of cosmic rays and radiation, but the 3D distributions of gas, the ISRF, and the cosmic rays are not well known.

\begin{figure*}
\begin{center}
\begin{tabular}{lll}
\includegraphics[height=0.25\textwidth]{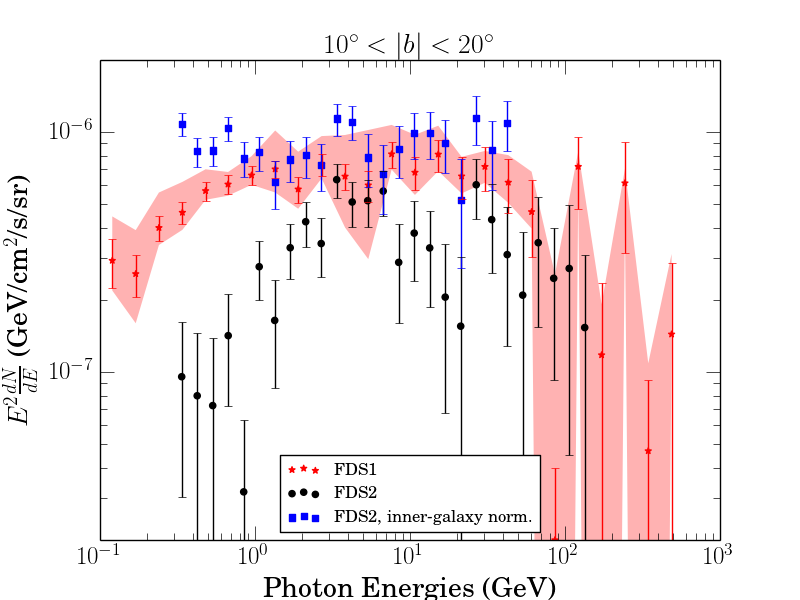} & \includegraphics[height=0.25\textwidth]{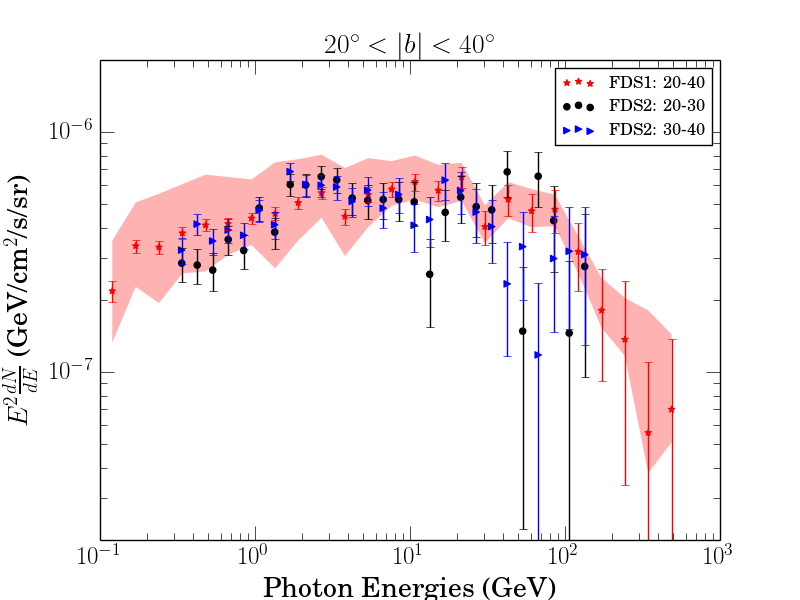} & \includegraphics[height=0.25\textwidth]{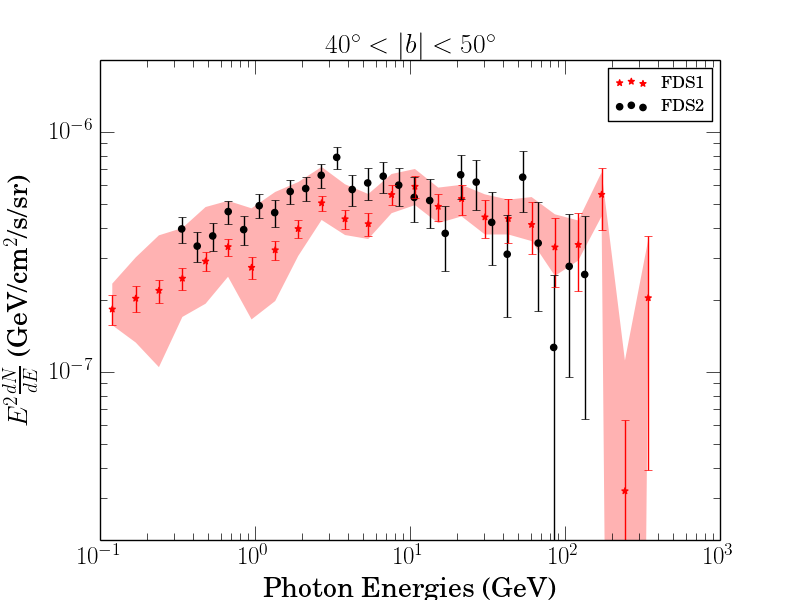}
\end{tabular}
\caption{The gamma-ray spectra for FDS1 and FDS2 (see text for definitions) in the different relevant latitude regions. Black and blue points indicate FDS2 spectra, red points indicate FDS1 spectra, with the red band describing the FDS1 systematic uncertainties. Since FDS1 has three bands, the middle two bands from FDS2 are plotted along with the second band of FDS1. Additionally in the 10-20$^\circ$ region we have included FDS2 data using the inner galaxy region of interest (see Appendix \ref{app:innergalaxy}), labelled `inner-galaxy norm'.}
\label{bubbles}
\end{center}
\end{figure*}

The \emph{Fermi} Collaboration \citep{apj2014_14077905} obtained an estimate of systematic uncertainties by varying the background model and the signal templates for the Bubbles. The results from the southern sky have noticeably smaller systematic uncertainties than the northern sky, especially at low energy and low latitude, most likely due to lower levels of dust and gas along the lines of sight toward the southern inner Galaxy. For the same reason, the southern sky is cleaner in the microwave data, which we will use to study the synchrotron emission. Accordingly, we use only the gamma-ray spectra extracted from the southern Bubble. 

We label these spectra \emph{Fermi} Data Set 1 (FDS1), and use them for our main fits, as the broad latitude bands allow for considerable statistical power. We estimate the effect of systematic errors by taking the difference between the central value and the edge of the systematic uncertainty band, treating these differences as upper and lower error bars, and adding them in quadrature to the upper and lower statistical error bars. (This procedure neglects correlations between the systematic uncertainties, but gives an estimate of the degree to which the bin-by-bin results can shift due to choice of background model.)

\subsubsection{Fermi Data Set 2: Matching the 2013 Latitude-Dependent Analysis of the Bubbles}

A more fine-grained latitude-dependent analysis was conducted independently of the \emph{Fermi} Collaboration in 2013 \citep{Hooper:2013rwa}; to test consistency with the results of that analysis, we use a very similar method and dataset to generate a secondary collection of gamma-ray spectra. We correct a bug in the smoothing of the diffuse background model for that paper, and impose cuts to remove events with poor angular resolution (following the procedure developed by \cite{Portillo:2014aaa}, as discussed by \cite{Daylan:2014rsa}; we remove half the total events), but otherwise proceed exactly as described there; in particular, we model the sky as a linear combination of the \emph{Fermi} \texttt{p6v11} diffuse model,\footnote{Available online at 
\texttt{http://fermi.gsfc.nasa.gov/ssc/data/}
\texttt{access/lat/BackgroundModels.html}} an isotropic background, the Bubbles sliced in 10-degree-wide latitude bands, and a dark-matter-motivated template corresponding to a contracted (inner power-law slope 1.3) Navarro-Frenk-White (NFW) profile, squared and projected along the line of sight (see \cite{Hooper:2013rwa} for details). This last component serves to absorb the GC excess emission, brightest at energies of 1-3 GeV (see e.g. \cite{Goodenough:2009gk,Hooper:2010mq,Boyarsky:2010dr,Hooper:2011ti,Abazajian:2012pn}), which is now known to extend into the inner Galaxy \citep{Hooper:2013rwa, Huang:2013pda, Daylan:2014rsa, 2015JCAP...03..038C}. We make use of the Pass 7 (V15) reprocessed data taken between August 4, 2008 and December 5, 2013, using only front-converting, Ultraclean class events which fall in the top two quartiles of the CTBCORE parameter, as described by \cite{Portillo:2014aaa}. We also apply standard cuts to ensure data quality (zenith angle $<100^{\circ}$, instrumental rocking angle $<52^{\circ}$, \texttt{DATA\_QUAL} = 1, \texttt{LAT\_CONFIG}=1). The fit is performed over the full sky and the bands at equal latitude north and south of the Galactic plane are required to share the same spectrum. There are 30 logarithmically spaced energy bins between 0.3 and 300 GeV.

We label the resulting spectra as \emph{Fermi} Data Set 2 (FDS2); we use them as a cross-check on the FDS1 results, and to allow direct comparison with previous results \citep{Hooper:2013rwa}.  We plot FDS1 and FDS2 together for each latitude region in Fig.~\ref{bubbles}. For latitudes greater than $20^\circ$, the consistency is very good.

\begin{figure*}
\begin{center}
\begin{tabular}{lll}
\includegraphics[height=0.25\textwidth]{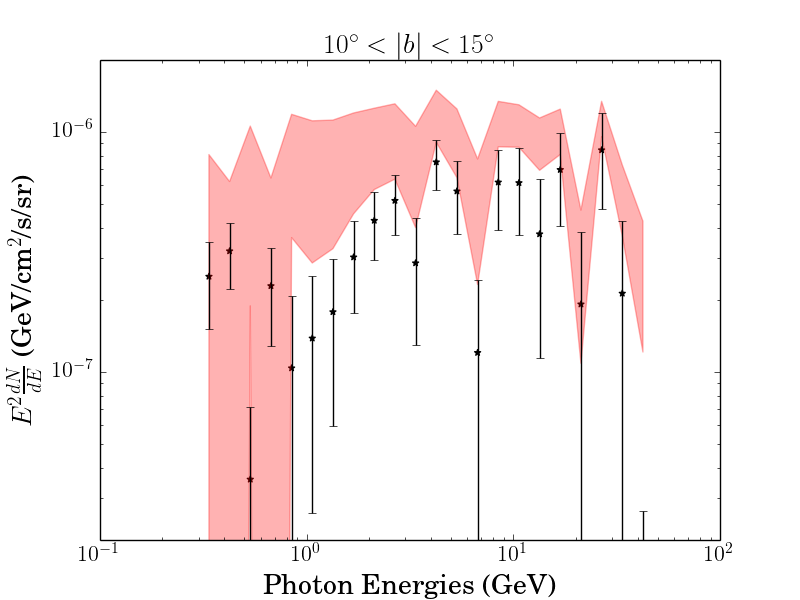} & \includegraphics[height=0.25\textwidth]{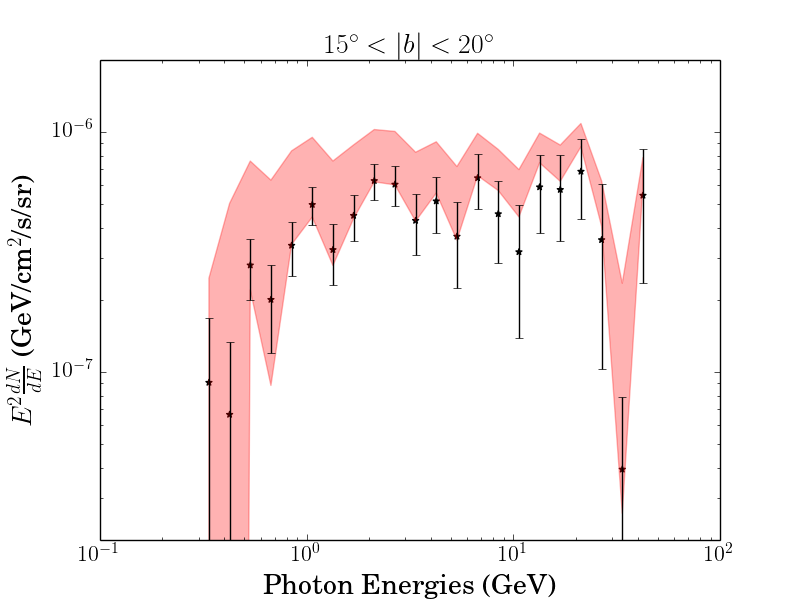} & \includegraphics[height=0.25\textwidth]{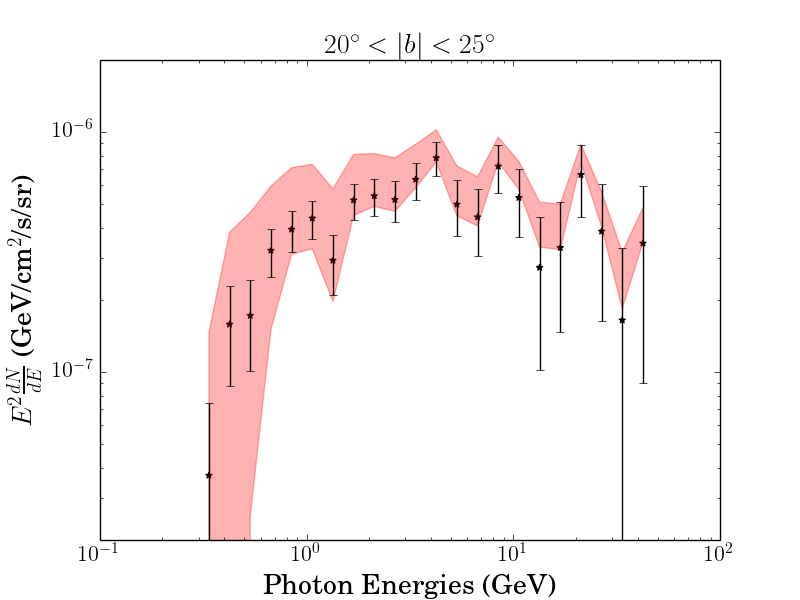}
\end{tabular}
\begin{tabular}{ll}
\includegraphics[height=0.25\textwidth]{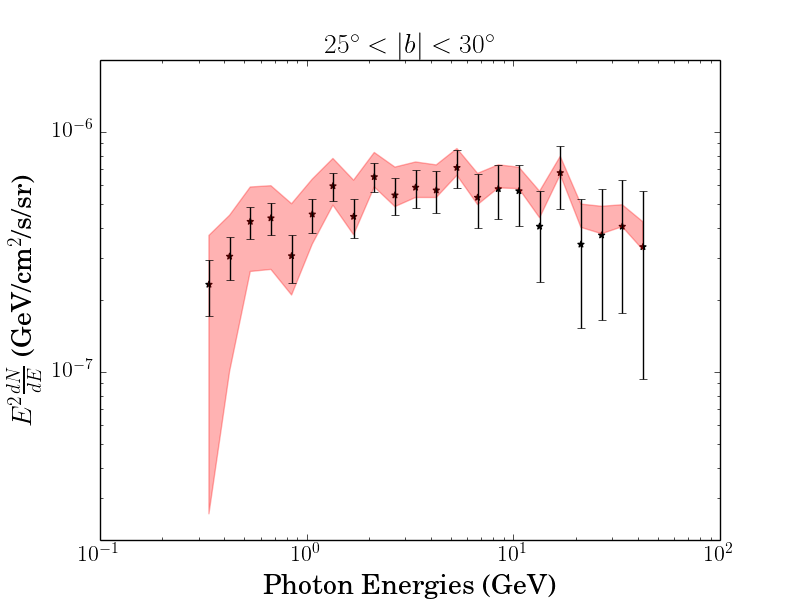} & \includegraphics[height=0.25\textwidth]{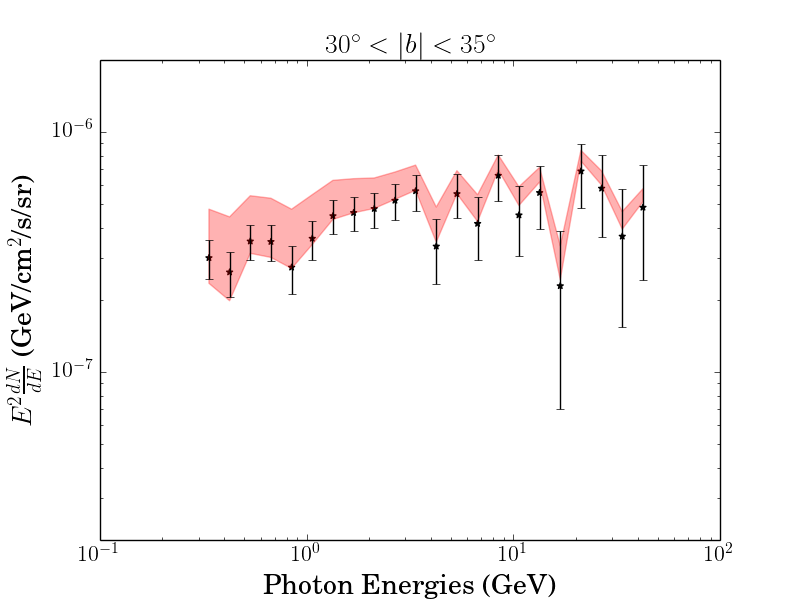} 
\end{tabular}
\caption{The gamma-ray spectra for FDS3, in each of five $5^\circ$-wide latitude bands. Black points indicate the spectra obtained using the \texttt{p6v11} diffuse model, with statistical error bars. Red bands give an estimate of the systematic uncertainty due to modeling of the diffuse background, with the minima and maxima of the band in each energy bin corresponding to fluxes one standard deviation away from the average of the spectra for 15 different diffuse models (including the default one).}
\label{fds3}
\end{center}
\end{figure*}

The FDS2 dataset does \emph{not} include an estimate of systematic uncertainties, since it is intended mostly as a cross-check with the literature. The lack of a systematic uncertainty band is a particularly acute problem in the lowest-latitude region we consider, $10^\circ < |b| < 20^\circ$, where the diffuse backgrounds are bright and their modeling can significantly affect the extracted spectrum for the Bubbles; for the default fitting procedure for FDS2, the results we obtain for this region at low energies appear quite different from FDS1, and in particular the spectrum is strongly suppressed at low energies. However, this is not a robust result -- in Fig.~\ref{bubbles}, we demonstrate the effect on the spectrum of performing the fit in a limited region of interest ($20^\circ \times 20^\circ$) rather than over the whole sky. This change to the analysis removes the suppression at low energies and produces results much more consistent with FDS1. Thus results from the $10-20^\circ$ band for FDS2 should be treated with caution. We will discuss this issue further in Appendix \ref{app:innergalaxy}.

\subsubsection{Fermi Data Set 3: A New Fine-Binned Analysis of the Bubbles}

Lastly, we conduct a new and more finely-binned analysis of the Bubbles spectrum employing the recently released Pass 8 \emph{Fermi} data, up to Mission Elapsed Time = 455067830 seconds (June 3, 2015). In order to facilitate comparison with the synchrotron data, we perform the fit over the southern sky only, masking latitudes $b > -10^\circ$, and divide the Bubbles into five latitude bins between $b = -10^\circ$ and $b=-35^\circ$ (angle south of the plane = $10-15^\circ$,  $15-20^\circ$, $20-25^\circ$, $25-30^\circ$, $30-35^\circ$), in addition to a `cap' template for the Bubbles with $b < -35^\circ$. We choose $35^\circ$ as our cutoff for the detailed study, because beyond this latitude there is little evidence for synchrotron emission.

We use data from the \texttt{ultracleanveto} event class to minimize cosmic ray contamination, and select the best quartile of events by their point spread function / angular resolution. We impose the recommended data quality cuts \texttt{(DATA\_QUAL$>$0)\&\&(LAT\_CONFIG$==$1)}, and a zenith angle cut of $90^\circ$. While we retain the energy binning of FDS2, we find that due to low statistics in our narrow latitude slices, the fit is numerically unstable at energies above 50 GeV. Accordingly, we show only results up to 50 GeV for this dataset.

By default we use the \texttt{p6v11} diffuse model for our background, as in FDS2; however, to estimate systematic uncertainties, we also test the effect of using 14 different models based on the public \texttt{GALPROP} code \citep{galprop, Strong:1999sv}. We test Models A and F-R as defined by \cite{2015JCAP...03..038C} (models F-R were originally taken from \cite{2012ApJ...750....3A}). In all cases we float the ICS and gas-correlated components of the model independently in each energy bin, and also allow a floating isotropic component. Despite the additional freedom to fit the ICS and gas-correlated emission separately in these models, we find that the  \texttt{p6v11} diffuse model provides a better overall fit to the data than any of the other 14; accordingly, we primarily employ the Bubbles spectra derived with the \texttt{p6v11} background model, and use the other spectra to estimate systematic uncertainties.

By default, we do not include an NFW-like component to absorb the GC excess discussed above, as the excess is not confidently detected at more than $10^\circ$ from the GC. We have tested the impact both of including an NFW-like component, and of changing the region of interest to include only $|l| < 20^\circ$ (similarly to the modified region of interest discussed for FDS2 above); in both cases, and for almost all diffuse models, the resulting perturbations to the spectrum lie almost entirely within the $1\sigma$ statistical error bars. (The one exception is Model A, where the deviations are outside the statistical error bars; since we only use the GALPROP models to estimate systematic uncertainties, we do not consider this isolated occurrence to be a problem.)

To approximately include the systematic errors due to uncertainties in the diffuse background, in each energy bin we take the standard deviation of the results for the different diffuse models, and treat this as a systematic error bar to be added in quadrature to the statistical errors. We refer to the resulting spectra as \emph{Fermi} Data Set 3 (FDS3); we show the average spectra, the standard deviation in each bin, and the data points for our default model in Fig.~\ref{fds3}.

\subsection{Microwave data from \emph{WMAP}}

For the microwave emission, we use data from the Wilkinson Microwave Anisotropy Probe (WMAP). Our main dataset is taken from \cite{2012ApJ...750...17D} and based on WMAP7 data. It describes the (unpolarized) microwave intensity associated with the haze/Bubbles for $-6^\circ > b > -90^\circ$ in degree-wide latitude bands, over 5 different frequencies: $(22.8, 33.2, 41.0, 61.4, 94.0)$ GHz. In each band, the emission is averaged over the longitude range $-5^\circ < l < 15^\circ$.

For comparison to the literature, we also include the results of \cite{2013A&A...554A.139P}: these are not as useful for our main analysis as they are not broken down by latitude, focusing on the latitude range $5$-$15$ degrees. These data were used by \cite{apj2014_14077905} to perform their comparison between the gamma-ray Bubbles and the microwave haze, and cover four frequencies: $(22.8, 28.5, 33.0, 40.9)$ GHz; the $28.5$ GHz band is taken from \emph{Planck} data, and the others from WMAP data.

\begin{figure}
  \begin{center}
    \includegraphics[height=0.35\textwidth]{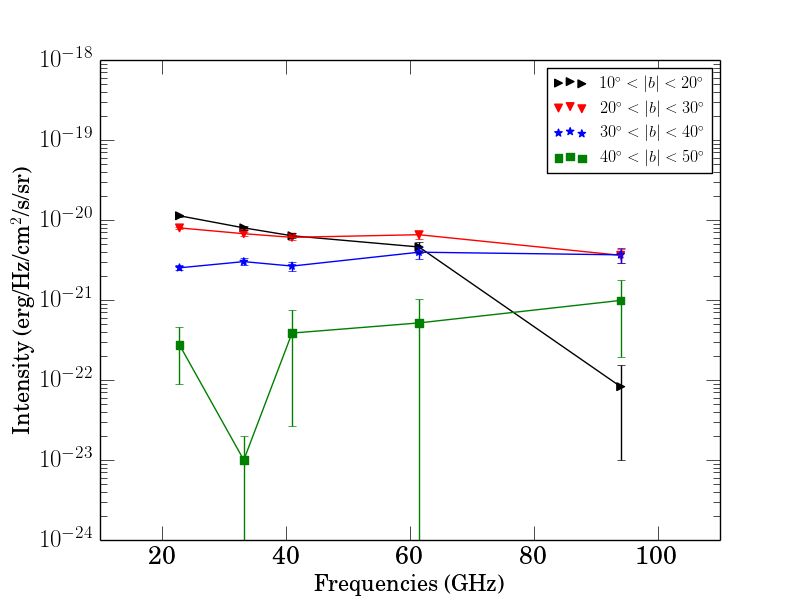}
    \caption{The microwave spectrum associated with the WMAP Haze \citep{2012ApJ...750...17D}, 
    using WMAP7 data, in each of four ten-degree latitude bands with $-5^\circ < l < 15^\circ$. The frequency bands are $(22.8, 33.2, 41.0, 61.4, 94.0)$ GHz.}
      \label{WMAPall}
  \end{center}
  \end{figure}

In order to obtain the average microwave spectrum in the latitude bins we were concerned with (3, 4 or 5 bins depending on the choice of FDS), we simply averaged over the 1-degree latitude bands contained in our larger bins. The provided errors were added in quadrature (to obtain the error on the sum), and then divided by the number of 1-degree bands in the bin (to obtain the error in the average). The microwave spectrum in ten-degree latitude bands is shown in Fig.~\ref{WMAPall}. (For comparison to FDS1 or FDS3 gamma-ray data, the bands are combined or subdivided accordingly.)

\cite{2012ApJ...750...17D} focused on the results from the lowest three frequency bands, as at higher frequencies the signal is fainter relative to the background emission and the systematic uncertainties become large. In this frequency regime, Dobler found consistency between the spectra of the \emph{Fermi} Bubbles and WMAP Haze, assuming a fully leptonic origin for both signals. We follow this approach, but for completeness, we show how the results change if all five frequencies are used in Appendix \ref{app:systematics}.

\subsection{Review of essential results for inverse Compton scattering and synchrotron radiation}

Under the hypothesis we examine in this article, the observed gamma rays (microwaves) are produced via ICS (synchrotron), from interactions between electrons and the ISRF (galactic magnetic field). In this subsection we provide a brief review of essential formulae for ICS and synchrotron radiation.

\subsubsection{Inverse Compton Scattering}

Consider an isotropic gas of high energy electrons inside the bubbles. Low energy photons from the ISRF scatter off these electrons to produce high energy gamma-rays. We concern ourselves with the quantity $\frac{dN_{\gamma, \epsilon}}{dtd\epsilon_1}$, the rate of production of scattered photons $N_{\gamma, \epsilon}$ per unit scattered photon energy $\epsilon_1$ per unit time, due to an electron with energy $\gamma m_e c^2$ scattering on a photon with energy $\epsilon$. This quantity is given by~\citep{1970RvMP...42..237B}:
\bea
\frac{dN_{\gamma,\epsilon}}{dtdE_1} & = & \frac{2\pi r_0^2mc^3}{\gamma}\frac{n(\epsilon)d\epsilon}{\epsilon}\cr
& \times & \left[2q\ln q+(1+2q)(1-q)+\frac{1}{2}\frac{(\Gamma_\epsilon q)^2}{1+\Gamma_\epsilon q}(1-q)\right],\cr
& &
\label{ICS_eq}
\eea
where $$\Gamma_\epsilon = \frac{4\epsilon\gamma}{mc^2}, \ \ q = \frac{E_1}{\Gamma_\epsilon(1-E_1)}.$$ Here $n(\epsilon)$ is the distribution of incoming soft photons and $E_1 = \frac{\epsilon_1}{\gamma m_e c^2}$ is the dimensionless ratio of the scattered photon energy to the original electron energy. Note that this is the scattering rate for a distribution of photons striking a single electron with energy $\gamma m_e c^2$; to obtain the full $\gamma$-ray spectrum, we must integrate over the electron spectrum. 

\subsubsection{Synchrotron Radiation}
Under the influence of a magnetic field, high energy electrons will undergo helical motion and produce synchrotron radiation. Consider an electron of energy $E=\gamma m_e c^2$, whose velocity makes an angle $\alpha$ with the magnetic field direction. The general expression for synchrotron emission due to an arbitrary electron distribution $N(\gamma,\alpha, r, t)$ is  given by~\citep{1970RvMP...42..237B}:
\bea
\frac{dW}{d\nu dt} & =  &\int d\Omega_\alpha \int d\gamma\frac{\sqrt{3}e^3B\sin\alpha}{mc^2}\frac{\nu}{\nu_c}\cr
& \times & N(\gamma,\alpha)\int_{\nu/\nu_c}^\infty d\xi K_{5/3}(\xi)
\label{syncheq}
\eea
where $\nu$ is the frequency of the observed synchrotron radiation, $B$ is the magnetic field, $K_{5/3}$ is the modified Bessel function of order $5/3$ and $$\nu_c = \frac{3eB\gamma^2}{4\pi mc}\sin\alpha$$ is the critical frequency. We assume an isotropic electron population, so the integral over pitch angle $\alpha$ evaluates to a constant. Note that if the electron spectrum $N(\gamma)\propto \gamma^{-p}$ (where $N(\gamma) = \frac{dN}{d\gamma} \propto \frac{dN}{dE}$), the resulting synchrotron spectrum has $\frac{dN}{dE} \propto E^{-\frac{p+1}{2}}$.

\subsection{The interstellar radiation field}

To calculate the ICS spectrum it is necessary to know the spatial and spectral distribution of the soft photon background, inside the \emph{Fermi} bubbles. In addition to the cosmic microwave background (CMB), which is an isotropic blackbody with temperature 2.725K, there are contributions from (re-scattered) starlight at (infrared) optical frequencies. An estimate for the spectrum of this interstellar radiation field (ISRF) is provided as part of the \texttt{GALPROP v54} distribution~\citep{Porter:2005qx}, sampled at 0, 0.1, 0.2, 0.5, 1, 2, 5, 10, 20, and 30 kpc above and below the Galactic plane. In our analysis the results for 1,2,5 and 10 kpc are most relevant, as the $10-50^\circ$ angular range we consider corresponds to $\sim 1.5-10$ kpc (for sources located directly above and below the GC). We performed a linear interpolation to estimate the ISRF spectrum at the middle of each latitude bin in our analysis, taking the distance from the Galactic plane to be given by $|8.5 \tan(b)|$ kpc, as appropriate for a structure centered at the GC. Here 8.5 kpc is the approximate distance between the Earth and the GC; this is the value assumed in \texttt{GALPROP v54}, consistent with recent determinations \citep{2012MNRAS.427..274S}.

In general, we hold the ISRF fixed; however, we will explore the effects of letting the normalization of the non-CMB contributions vary (while holding the CMB component fixed) in Appendix \ref{app:isrf}.

\subsection{Modeling the electron spectrum}
\label{sec:isrf}
 
Given the substantial statistical and systematic uncertainties in the gamma-ray and microwave data, we wish to avoid overly complex models for the underlying electron spectrum. Accordingly, we consider power-law spectra, which may:
\begin{enumerate}[(a)]
\item be featureless, $N(E) \propto E^{-p}$ (2 parameters, slope + normalization),
\item have a sharp cutoff at some energy, $N(E) \propto E^{-p}$ for $E < E_\mathrm{cut}$ (3 parameters, with the third being the cutoff energy),
\item have a smooth exponential cutoff at some energy, $N(E) \propto E^{-p}e^{-E/E_\mathrm{cut}}$ (3 parameters, with the third being the cutoff energy).
\end{enumerate}

In general, we will find that a simple power law is sufficient to fit the microwave data, whereas some cutoff is required to fit the gamma rays, if the high-energy data are included, consistent with the literature (see e.g. \cite{Finkbeiner:2003im, apj2014_14077905}). A sharp cutoff \citep{Su:2010qj} and an exponential cutoff \citep{apj2014_14077905} have both been employed in the literature, so we examine both, although the latter seems more physical. The effect of electrons above the cutoff in the exponential-cutoff case is small, but not negligible. 

One might also consider the case of a power law that breaks to a different power law at some energy, $N(E) \propto E^{-p_1}$ for $E < E_\mathrm{b}$, $N(E) \propto E^{-p_2}$ for $E > E_\mathrm{b}$ (4 parameters, with the third and fourth being the break energy and the slope above the break). We find that such broken power laws do not generally significantly improve the fit over power laws with cutoffs, and in the microwave or gamma-ray data taken individually, it becomes difficult to constrain all the parameters. For example, we fit a broken power law to the 20-40$^\circ$ FDS1 data, and obtained a minimum $\chi^2$ of 5.18 compared to 4.57 when fitting a single power law. In the case of a double power law a model with $p_1=2.60, p_2 \ge 5.00, E_b=2.03\TeV$ was favored; since the slope above the break is very steep, this model is quite similar to a single power law with a cutoff. Accordingly, we will not generally consider broken power law models, since they do not seem to be necessary to explain the data.

We note that in general, for reasonable magnetic fields, the microwave data probe lower electron energies than the gamma-ray data. For $\sin\alpha=1$, the critical frequency for synchrotron is given by $\nu_c = \frac{3 e B \gamma^2}{4 \pi m_e} \approx 42 \, \mathrm{Hz} \left(\frac{B}{10 \mu\mathrm{G} } \right) \gamma^2 $. Thus frequencies of 23-94 GHz correspond to $\gamma \approx 2-5 \times 10^{4} \times \left(\frac{B}{10 \mu\mathrm{G} } \right)^{-1/2} $, i.e. $E \approx 12-24$ GeV $\times  \left(\frac{B}{10 \mu\mathrm{G} } \right)^{-1/2}$. In contrast, the average energy of inverse-Compton scattered photons is $E \sim (4/3) \gamma^2 E_0$, where $E_0$ is the energy of the soft photon. Thus for a starlight spectrum with energy peaked around $\sim 1$ eV, the $\sim 100$ GeV gamma-rays visible in the \emph{Fermi} Bubbles correspond to electrons with at least $\gamma \approx 10^{5.5}$, i.e. $E \approx 150$ GeV. 1 GeV gamma-rays from ICS on starlight probe a similar energy range to the synchrotron spectrum (a few tens of GeV), but 1 GeV gamma-rays can also be produced by much higher-energy electrons scattering on the lower-energy photons of the CMB.

Accordingly, it is not at all surprising to find evidence for a high-energy break in the Bubbles (as already discussed by \cite{apj2014_14077905}) and no sign of such a break in the microwave spectrum.

\subsection{Fitting to the data}

In order to find the ICS spectrum produced by these electron spectra, we evaluate Equation \ref{ICS_eq} at each electron energy $E = \gamma m_e c^2$, scale by the number of electrons at that energy, $N(E)$, and then integrate over all electron energies $E$. The resulting spectrum of scattered photons can be compared to the gamma ray spectrum of the \emph{Fermi} Bubbles. We perform a simple $\chi^2$ minimization, with error estimates as described above (statistical+systematic for FDS1 and FDS3, statistical only for FDS2). As our test statistic, we use $\Delta \chi^2$ relative to the global $\chi^2$ minimum. For example, if there is one additional degree of freedom in the model relative to the best-fit point, the region with $\Delta\chi^2 <0.989$ corresponds to the 68 percent confidence interval.

When fitting to the microwave data, we similarly use Equation \ref{syncheq} to determine the expected microwave flux, and include the strength of the magnetic field as an additional free parameter in the fit. When fitting simultaneously to the microwave and gamma-ray data, we scan the magnetic field for each possible set of parameters for the electron spectrum, and thus construct a combined $\chi^2$ value:

\begin{align} \chi^2_\mathrm{total}(p, E_\mathrm{cut}, B) & = \chi^2_\mathrm{ICS}(p, E_\mathrm{cut}) \nonumber \\
& +\chi^2_\mathrm{synch}(p, E_\mathrm{cut}, B). \end{align}

When we marginalize over specific model parameters, to obtain constraints on the other parameters, we employ the profile likelihood method \citep{Rolke:2004mj}.

\section{Comparison of `Overall' Gamma-Ray and Microwave Spectra}
\label{sec:overallcomp}

We begin by re-checking the consistency between FDS1 and the microwave spectrum presented by \cite{2013A&A...554A.139P}. In particular, we focus on the FDS1 spectrum in the $40-60^\circ$ band, as at high latitudes the spectrum of the Bubbles is quite stable as a function of latitude. Following \cite{apj2014_14077905}, we take the ISRF to be given by its value 5 kpc away from the Galactic plane, directly south of the GC.

\subsection{Individual fits}

Fitting to the gamma-ray data alone, for a power law with an exponential cutoff, we find best-fit parameters of $p = 2.19^{+0.12}_{-0.30}, \ E_\mathrm{cut} = 1.62^{+0.42}_{-0.60} \TeV$.

Fitting to the microwave data alone, if we model the electron spectrum as a power law with a cutoff, we find a best-fit power-law slope of $p=2.13$ (and $p < 2.49$ at 68\% confidence). However, since this model has four parameters (magnetic field, power-law slope, cutoff and amplitude) and there are only four data points, covering a relatively small energy range, we cannot set sensible limits on the cutoff energy or a lower limit on the power-law slope. A very hard power-law can be compensated by a relatively low cutoff, with the magnetic field tuned to ensure that the resulting sharply peaked synchrotron spectrum falls within the range of the \emph{WMAP} data. Given the relatively high cutoff found in the gamma-ray analysis, and the fact that the microwave data are not sensitive to such a high cutoff, we instead fit the microwave data with a simple power law. In this case the effect of the magnetic field and amplitude parameters are perfectly degenerate, so we cannot constrain either of these separately, but we can set a limit on the power-law slope. We find $p=2.13^{+0.36}_{-0.33}$. 

\subsection{Combined fit}

\begin{figure*}
  \begin{center}
  \begin{tabular}{ll}
    \includegraphics[height=0.35\textwidth]{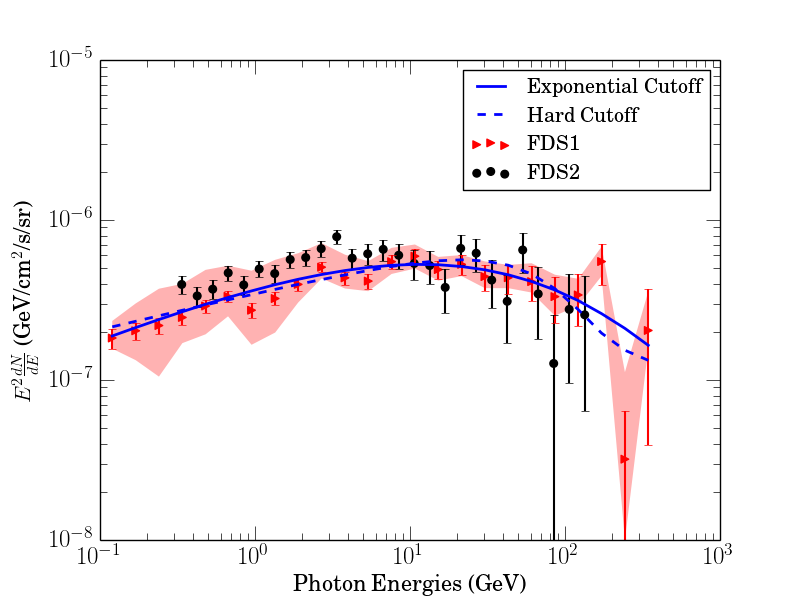} & \includegraphics[height=0.35\textwidth]{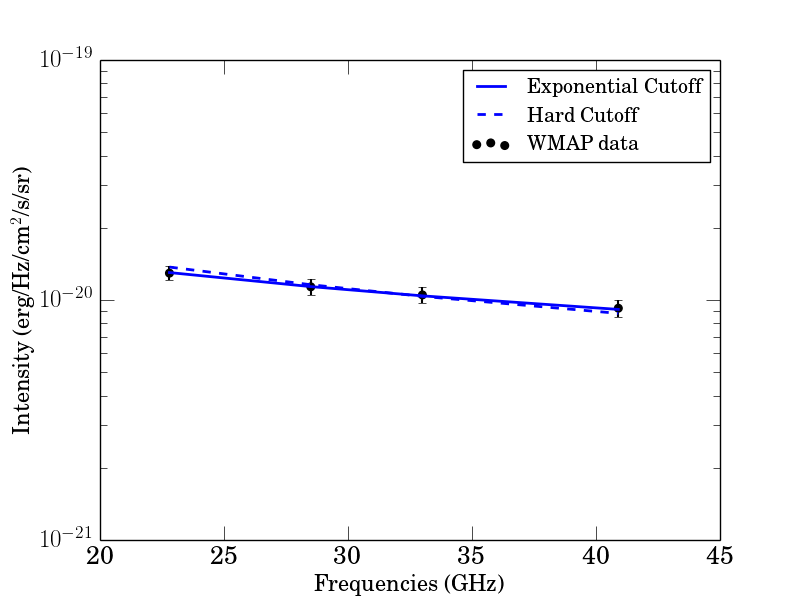}
    \end{tabular}
    \caption{In the left (right) panel we plot the combined best-fit models against the high-latitude gamma-ray ($-b=5-15^\circ$ microwave) data. In the left panel, the triangular data points represent FDS1 in the 40-60$^\circ$ band, with error bars denoting statistical errors only; systematic errors are shown by the red band. The circular data points represent FDS2 in the 40-50$^\circ$ band.  In the right panel, circular points describe the microwave data; errors are statistical only. In both panels, the best-fit models to FDS1 + microwave data, with an exponential high-energy cutoff and hard high-energy cutoff, are shown by the solid and dashed lines respectively.}
      \label{ICSsynch_plot}
  \end{center}
\end{figure*}

As discussed previously by e.g. ~\cite{Dobler:2009xz, Su:2010qj, 2012ApJ...750...17D, apj2014_14077905}, the consistency in the required power-law slopes suggests that the WMAP Haze and the \emph{Fermi} Bubbles spectrum can be described by the same distribution of high energy electrons.

We now construct the combined $\chi^2$ as discussed previously, for a power-law electron spectrum with an exponential cutoff, under the hypothesis that both signals arise from the same electron population. Marginalizing over all but one of the model parameters, using the profile likelihood method, we obtain the constraints on each of the model parameters in turn. We focus here on the slope and cutoff for the electron spectrum, and the magnetic field, rather than the absolute value of the overall amplitude (which holds information on the electron density inside the Bubbles, but requires a spatial model of the Bubbles for a detailed interpretation).

Using FDS1 and a power-law model with exponential cutoff for the electron spectrum, we find preferred parameters:

\begin{equation} B = 8.32^{+3.06}_{-0.86} \ \mu\mathrm{G}, \quad p = 2.19^{+0.10}_{-0.26}, \quad E_\mathrm{cut} = 1.62^{+0.37}_{-0.62} \TeV.\end{equation}

\noindent with $\chi^2$/dof = 0.27. If the exponential cutoff is replaced by a hard cutoff, we find instead:

\begin{equation} B = 6.31^{+0.98}_{-0.55} \ \mu\mathrm{G}, \quad p = 2.52^{+0.06}_{-0.06}, \quad E_\mathrm{cut} = 3.22^{+0.80}_{-0.93} \TeV.\end{equation}

\noindent with $\chi^2$/dof = 0.41. Switching to the FDS2 gamma-ray dataset, and using the 40-50$^\circ$ latitude band, we find for an exponential cutoff:
\begin{equation} B = 9.55^{+4.08}_{-2.02} \ \mu\mathrm{G}, \ p = 1.89^{+0.20}_{-0.21}, \ E_\mathrm{cut} = 0.810^{+0.280}_{-0.210} \TeV, \end{equation}
\noindent with $\chi^2$/dof = 0.72 and for a hard cutoff,
\begin{equation}B = 5.89^{+0.31}_{-0.50} \ \mu\mathrm{G} , \ p = 2.43^{+0.03}_{-0.01}, \ E_\mathrm{cut} = 1.62^{+0.33}_{-0.26} \TeV\end{equation} with $\chi^2$/dof = 0.82. In all cases the fit is very good, although the low $\chi^2$/dof in the case of FDS1 likely reflects generous estimates of systematic errors (and neglect of bin-to-bin correlations).

We see that the modeling of the high-energy cutoff can have an impact on the best-fit parameters, although the differences are comparable to the uncertainties. Models with smoother exponential cutoffs tend to prefer somewhat higher magnetic fields, lower cutoff energies, and harder spectra below the cutoff. Since an exponential cutoff is more physically reasonable than an abrupt break, we will use exponential cutoffs by default in the remainder of this article. Similarly, the FDS2 dataset prefers slightly harder spectra and lower cutoff energies than FDS1, although the results are consistent within the uncertainties.

We show the ICS and synchrotron spectra resulting from the models that provide the best fits to the FDS1 data in Fig.~\ref{ICSsynch_plot}.

A very similar analysis was done by \cite{apj2014_14077905} , assuming a power law with an exponential cutoff for the electron spectrum, and resulting in the following constraints:
\bea
B & = & 8.4\pm 0.2 [\rm{stat}]^{+11.2}_{-3.5}[\rm{syst}] \ \mu\mathrm{G} \cr
p & = & 2.17\pm 0.05 [\rm{stat}]^{+0.33}_{-0.89}[\rm{syst}]\cr
E_\mathrm{cut} & = & 1.25\pm 0.13 [\rm{stat}]^{+1.73}_{-0.68}[\rm{syst}] \TeV.
\eea

All our results fall well within this uncertainty band, and our best-fit results for FDS1 and the exponential cutoff model agree closely, as expected since the inputs are nearly identical (they differ in the treatment of uncertainties). The smaller size of our error bars is almost certainly due to our approximate treatment of the systematic errors (adding in quadrature), which should accordingly be treated with some caution.

\section{A Latitude-Dependent Comparison}
\label{sec:latdep}

\subsection{Consistency of FDS1 and FDS2 with microwave data by latitude}
\label{sec:latresults}

\begin{figure*}
\begin{center}
\begin{tabular}{ll}
\includegraphics[height=0.4\textwidth]{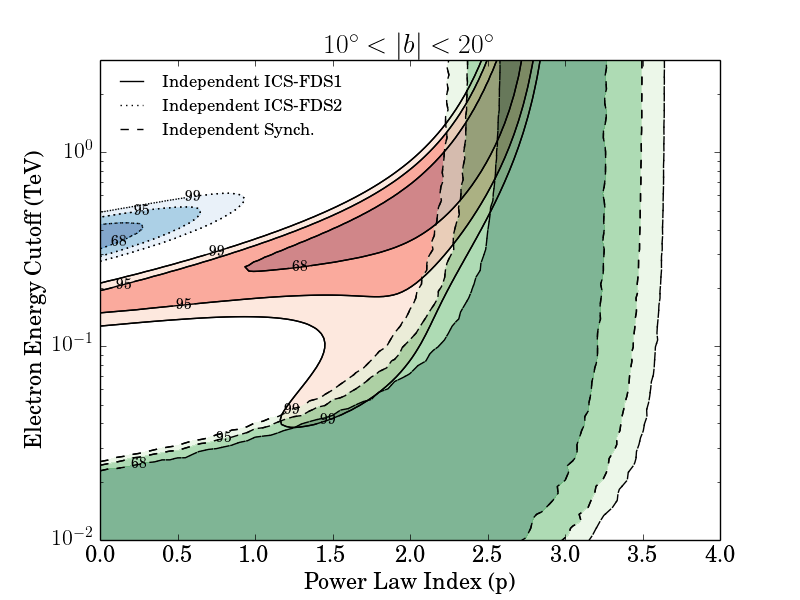} & \includegraphics[height=0.4\textwidth]{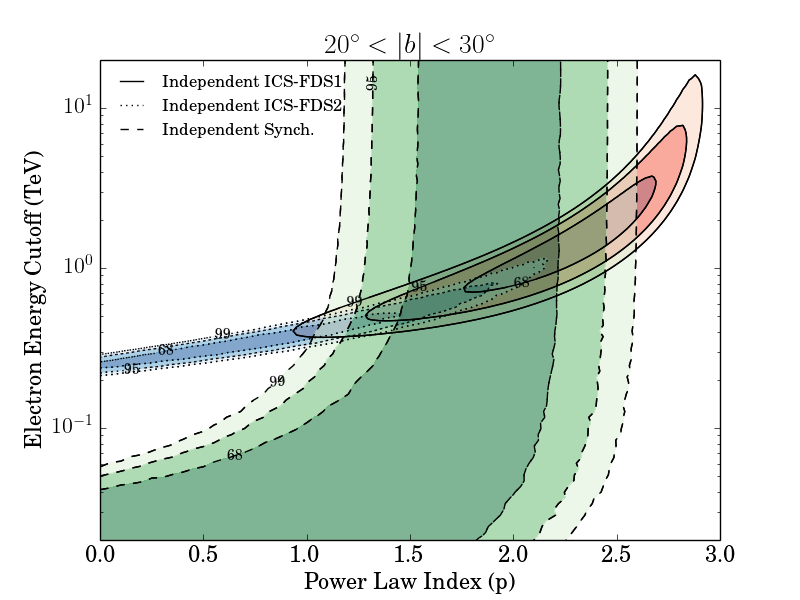} \\ 
\includegraphics[height=0.4\textwidth]{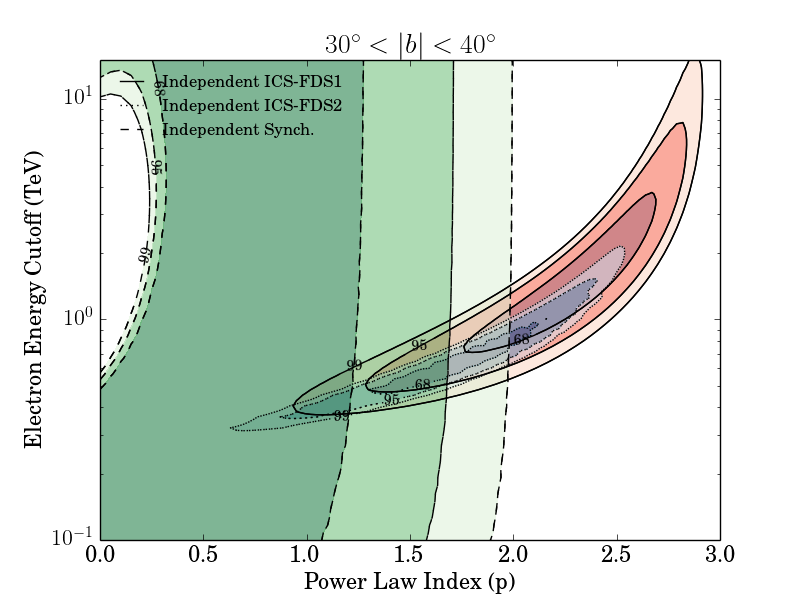} & \includegraphics[height=0.4\textwidth]{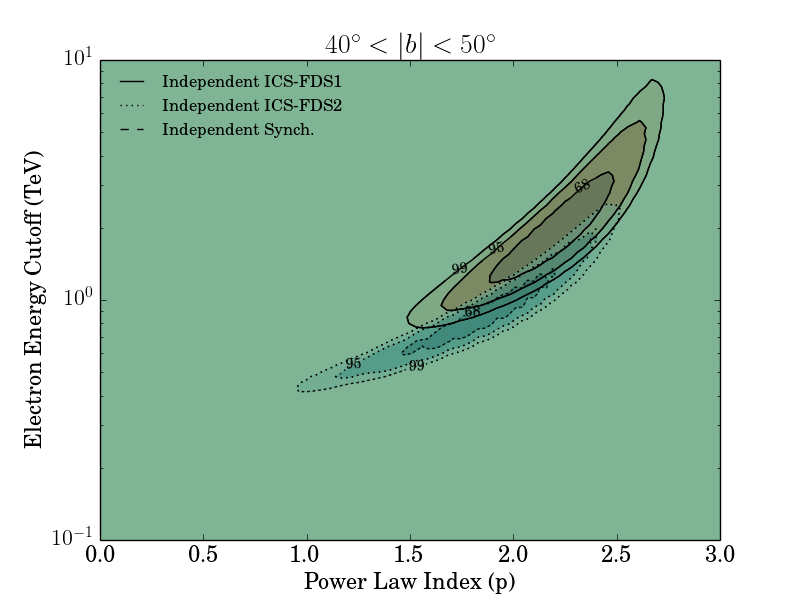} \\ 
\end{tabular}
\caption{$\Delta \chi^2$ contours for independent fits of a leptonic model to the FDS1, FDS2 (labelled `Independent ICS') and microwave data (labelled `Independent Synch'), where in both cases the electron spectrum is modelled as a power law with an exponential cutoff. We marginalize over the electron density, and the magnetic field in the case of the microwave data. The solid red (dotted blue) contours represent the fits to the FDS1 (FDS2) gamma-ray datasets; the dashed green contours represent the fit to the microwave data. In all cases the contours mark 68,95,99 percent confidence regions.
}
\label{indcont}
\end{center}
\end{figure*}

We now consider the consistency between the gamma-rays and microwaves in different latitude bands, in the context of our simple models for the electron spectrum. A similar analysis was done by \cite{Hooper:2013rwa}, but relied on reconstructing the electron spectrum from the gamma rays and then comparing to the microwave data, which complicated the propagation of uncertainties. As discussed above, we use latitude-binned microwave data from WMAP7 \citep{2012ApJ...750...17D}, and the latitude-binned gamma-ray datasets FDS1-3. For the gamma rays we consider a power law electron spectrum with an exponential cutoff; for the reasons discussed above, for the microwaves we consider only an unbroken power law electron spectrum. We take the ISRF model from \texttt{GALPROP} as described in Section \ref{sec:isrf}, and assume the ISRF at the center of each relevant latitude bin is a good approximation to the appropriately weighted ISRF integrated along the line of sight through the Bubbles. At this stage we do not include uncertainties in the ISRF model; we perform an initial analysis of the effect of varying this model in Appendix \ref{app:isrf}.

Taking the FDS1 gamma-ray dataset we have three latitude bins: $10-20^\circ$, $20-40^\circ$ and $40-60^\circ$. Only the first two can be used for a direct comparison with the microwaves, as in the $40-60^\circ$ bin we find the microwave emission is consistent with zero, so the spectrum cannot be constrained (although the absence of a signal can set upper bounds on the magnetic field, under the hypothesis of a leptonic origin for the gamma rays). This is consistent with the cutoff at $|b| \sim 35^\circ$ found by \cite{2012ApJ...750...17D}. As discussed above, when the electron spectral model is a simple power law, the microwave data cannot be used on their own to separately constrain the magnetic field and the amplitude of the electron spectrum.

In the $10-20^\circ$ latitude band we find a best-fit power law of $p = 2.90^{+0.25}_{-0.20}$ based on the microwaves (modeling the electron spectrum as a simple power law), whereas the gamma rays prefer $p=2.43^{+0.24}_{-0.75}$, with a high-energy (exponential) cutoff of $E_\mathrm{cut} = 1.19^{+2.31}_{-0.81}$ TeV. In the $20-40^\circ$ latitude band the fit to the microwaves yields $p = 1.55^{+0.15}_{-0.20}$, whereas the fit to the gamma rays gives  $p = 2.34^{+0.24}_{-0.33}$ and $E_\mathrm{cut} = 1.58^{+1.05}_{-0.63}$ TeV. (In the highest-latitude band, we find $p = 2.19^{+0.18}_{-0.18}$ and $E_\mathrm{cut} = 1.87^{+0.76}_{-0.46}$ TeV, fitting to the gamma rays alone.) A similar comparison is done using a a hard cutoff model for the electron spectrum in Appendix~\ref{app:hardcutoff}.

\begin{figure*}
  \begin{center}
    \includegraphics[width=0.45\textwidth]{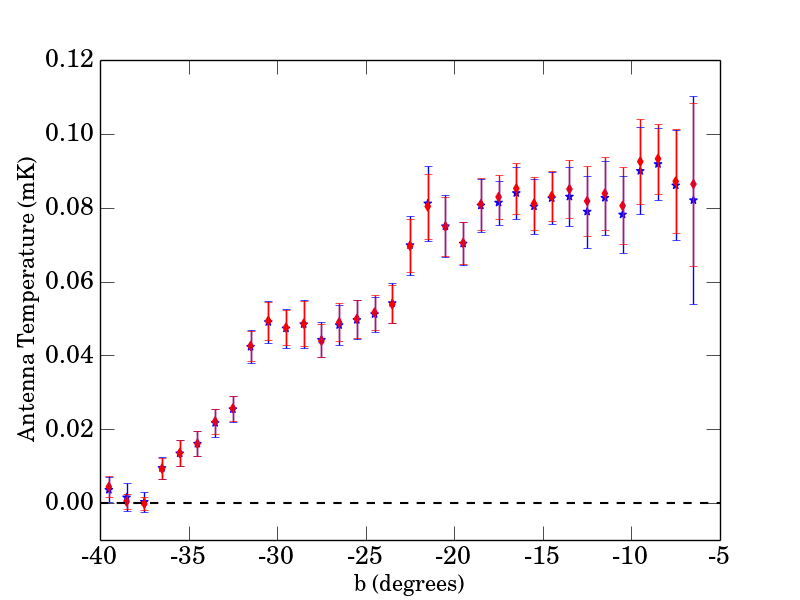}
    \includegraphics[width=0.45\textwidth]{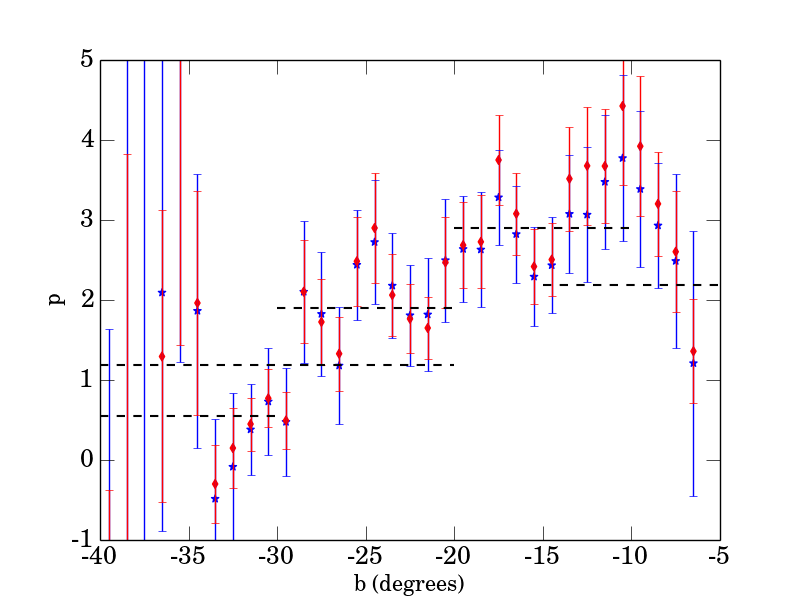}
    \caption{The left panel shows the amplitude of the microwave haze emission for the best-fit model at 23 GHz, as obtained by a power-law fit to the data; the horizontal dotted line indicates zero amplitude. The right panel shows the inferred slope of the power-law electron spectrum; the horizontal dashed lines indicate the slopes extracted by fits in the $|b|=10-20^\circ$, $20-30^\circ$, $30-40^\circ$, $20-40^\circ$ and $5-15^\circ$ regions of the southern sky, as discussed in Sections \ref{sec:overallcomp}-\ref{sec:latdep}. Blue stars (and error bars) indicate that the fit was performed using only the three lowest frequency bands of the microwave data; red diamonds (and error bars) were based on the fit using all five bands.
    }
      \label{fig:latsynch}
  \end{center}
  \end{figure*}
  
  \begin{figure}
  \begin{center}
    \includegraphics[width=0.45\textwidth]{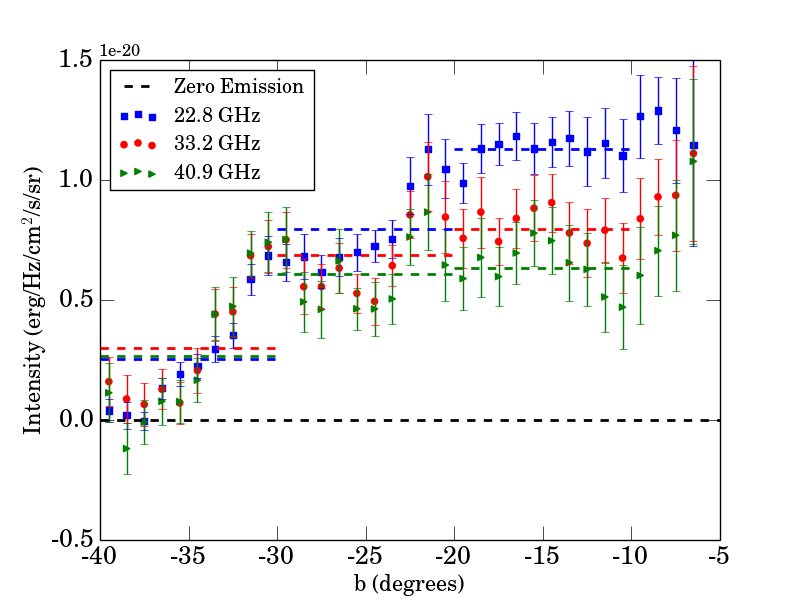}
    \caption{The microwave haze emission for the first three microwave frequency bands, represented by blue squares, red circles and green triangles, in 1-degree latitude bins. The dashed lines, of appropriate color, show the average values of microwave emission at these frequencies in the first three latitude bins used in our analysis: $|b| = 10-20^\circ, 20-30^\circ, 30-40^\circ$.}
      \label{fig:latdata}
  \end{center}
  \end{figure}

This comparison suggests some initial evidence for hardening in the spectrum required to explain the microwave data, as one moves to higher latitudes; in contrast, the gamma-ray data show no strong preference for a change in spectral slope with latitude.  One can perform a similar analysis for the FDS2 dataset, breaking the $|b|=20-40^\circ$ band into two separate bins. In the $|b|=20-30^\circ$ and $|b|=30-40^\circ$ latitude bands, the power-law electron spectra extracted from the microwave data are $p=1.90^{+0.20}_{-0.20}$ and $p=0.55^{+0.45}_{-0.54}$ respectively, again suggesting a hardening in the electron spectrum as distance from the Galactic plane increases, until $|b| \gtrsim 35^\circ$ where the microwave haze is not observed. In the gamma rays, the FDS1 and FDS2 results appear consistent except in the $10-20^\circ$ latitude band; as we have discussed, the FDS2 results are likely less reliable in this region. 

The $\Delta \chi^2$ contours for both FDS1 and FDS2 are plotted in Fig.~\ref{indcont}; here we also show the full $\Delta \chi^2$ contours as a function of power law slope and cutoff, for fits to the microwave spectrum, demonstrating the degeneracy between the slope and the cutoff discussed above. The $68,95,99$ percent confidence contours correspond to $\Delta\chi^2 = (2.28, 5.99, 9.21)$, respectively. The gamma-ray data exclude a maximum-energy cutoff below several hundred GeV, in the hypothesis where both signals originate from the same electrons, and the power law preferred by the microwave data is nearly independent of the cutoff once the cutoff energy becomes sufficiently high; this justifies our choice to model the electron spectrum as an unbroken power law when comparing our models to the microwave data alone. 

To explore the possible mid-latitude discrepancy in greater detail, we turn to the FDS3 dataset, with finer binning in latitude, and to the detailed WMAP data. 

\subsection{Detailed latitude variation of the electron spectrum inferred from microwaves}

If we model the electron spectrum as an unbroken power law $dN/dE \propto E^{-p}$ in fitting to the microwave data, there is no need for careful calculation of the synchrotron emission: it will simply be a power law with slope $(p+1)/2$, independent of the magnetic field strength, as mentioned in Section \ref{sec:methods}. Thus we can fit a power law directly to the microwave data, and translate the limits on its index into limits on $p$.

Starting with the microwave spectra obtained by \cite{2012ApJ...750...17D}, we perform a $\chi^2$ fit to a simple power law in each of the degree-wide latitude bands. In Fig.~\ref{fig:latsynch} we show the best-fit amplitude at 23 GHz as a function of latitude, and the best-fit value of $p$, with corresponding statistical error bars. As expected, the uncertainties on the power-law slope become very large for $b < -35^\circ$, where the amplitude of the microwave haze goes to zero. We also indicate in Fig.~ \ref{fig:latsynch} the previously stated best-fit values for the electron spectrum slope based on the microwave data in the $5-15^\circ$, $10-20^\circ$ and $20-40^\circ$ regions; we can see that (as expected) these results look broadly consistent with an average over the latitude bands in question. There is indeed a noticeable trend in the preferred slope, with the spectrum softening moving outward from $|b|=5-10^\circ$, but then a marked hardening trend from $\sim10^\circ$ out to $\sim 35^\circ$; this result holds whether the fit is performed over only the first three frequency bands or all five. It is possible that this hardening is due to systematic errors in the microwave data, but there does seem to be a consistent trend with latitude. In Fig.~\ref{fig:latdata} we show the raw data used in this fit, for the first three energy bins, and again the high-latitude hardening is manifest; the ratio of the low-energy to the high-energy data is noticeably larger at low latitudes. It is worth noting, however, that in all three energy bands the emission seems to approach zero around the same latitude ($b\sim -35^\circ$); the hardening does \emph{not} seem to reflect the high-energy emission extending to latitudes where the low-energy emission has already cut off.

  \begin{figure}
  \begin{center}
    \includegraphics[width=0.45\textwidth]{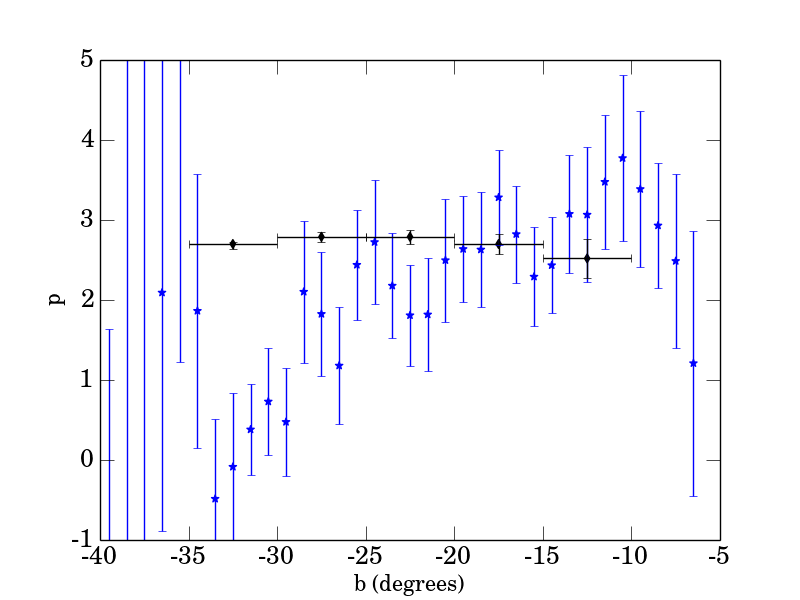}
    \caption{The inferred slope of the power-law electron spectrum from microwave data (blue stars) and gamma-ray data FDS3 (black diamonds), assuming an unbroken power law in both cases. We employ only the first three WMAP frequency bins, for the microwave data.}
      \label{fig:latcompare}
  \end{center}
  \end{figure}

\begin{figure*}
\begin{center}
\begin{tabular}{lll}
\includegraphics[height=0.25\textwidth]{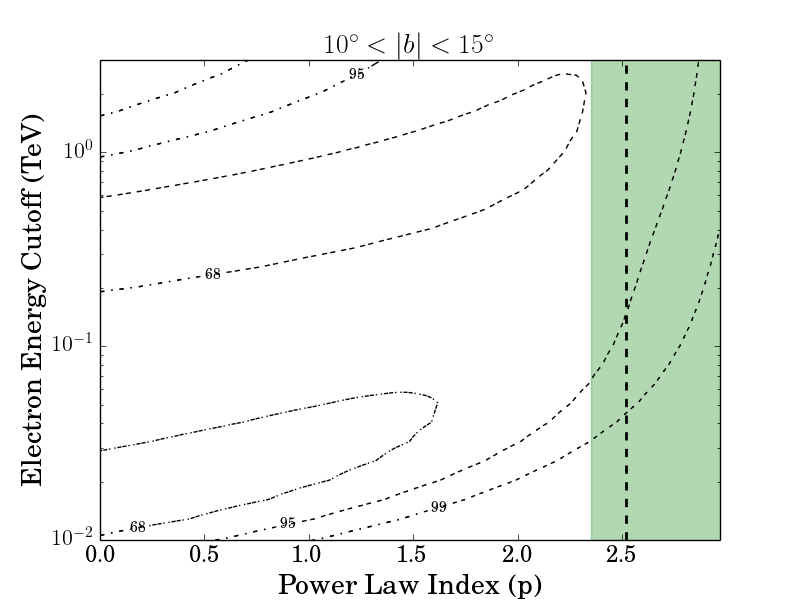} & \includegraphics[height=0.25\textwidth]{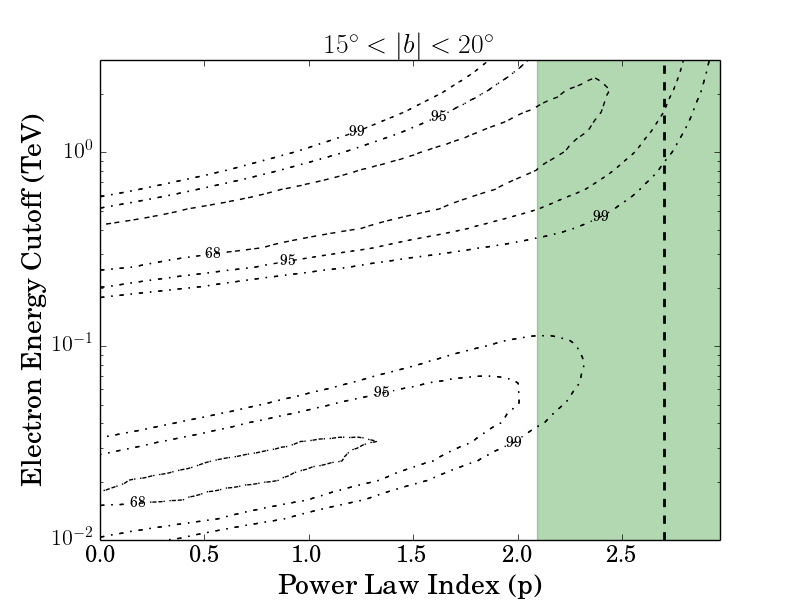} & 
\includegraphics[height=0.25\textwidth]{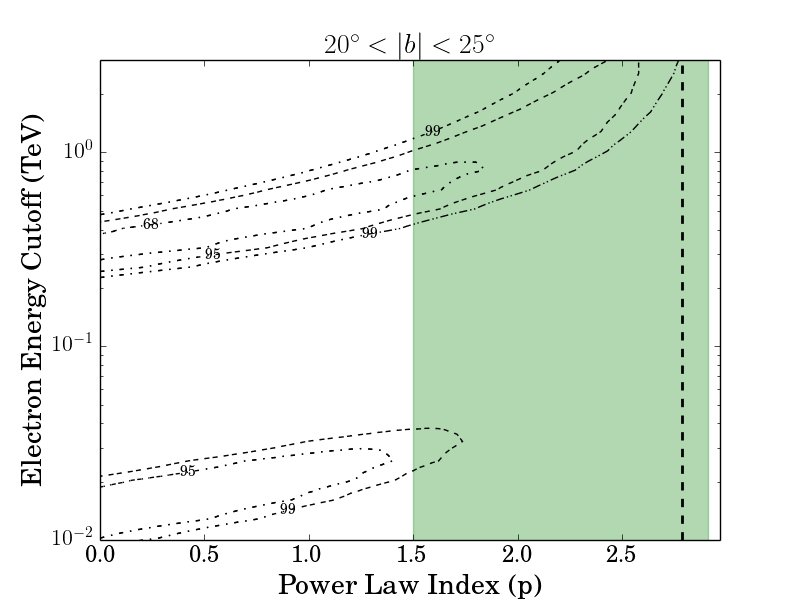}
\end{tabular}
\begin{tabular}{ll}
 \includegraphics[height=0.25\textwidth]{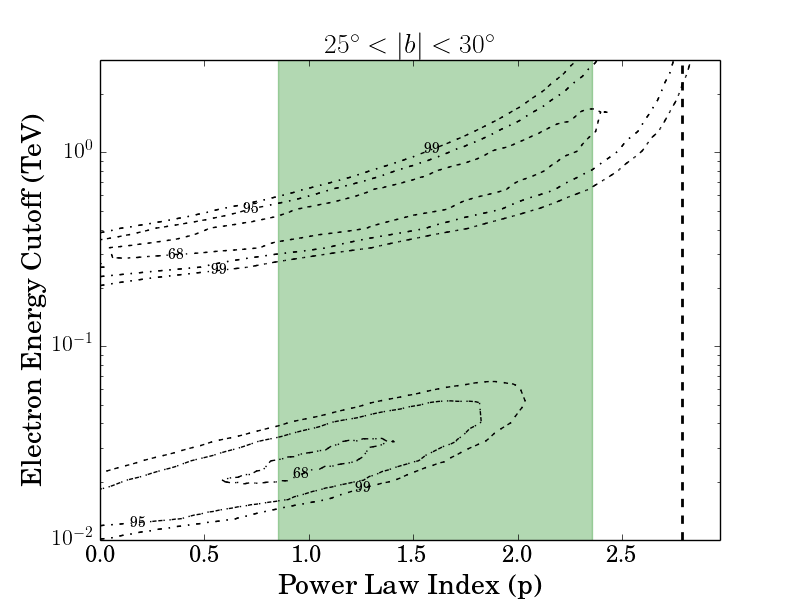} & 
\includegraphics[height=0.25\textwidth]{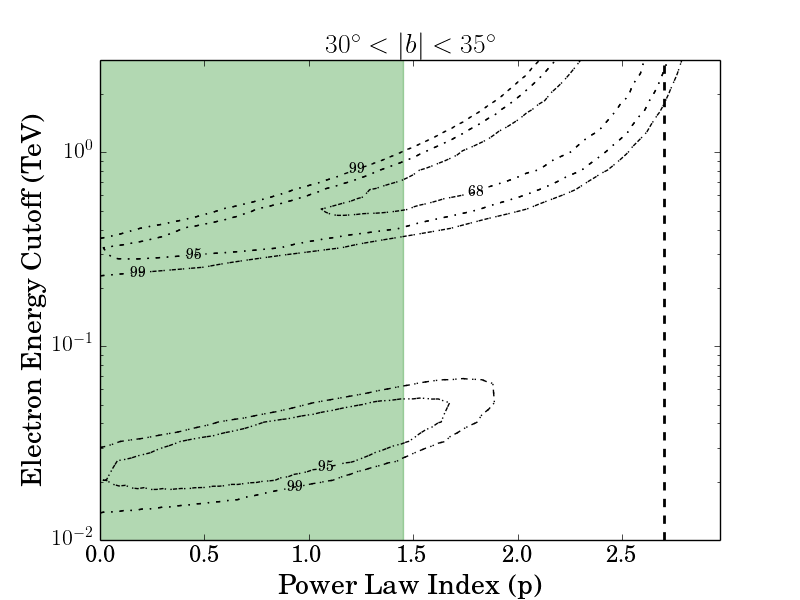} 
\end{tabular}
\caption{We show the 68,95,99 percent confidence contours for fits to FDS3 alone, marginalizing over amplitude, in each of the five degree latitude bins ($-b = 10-15^\circ, 15-20^\circ, 20-25^\circ, 25-30^\circ, 30-35^\circ$). The electron spectrum was modelled as a power law with exponential cutoff. The green bands represent the predicted slope as in the right panel of Fig.~\ref{fig:latsynch}; the central value is the average of the data points within the appropriate latitude range, and the width is determined by the average of the $1\sigma$ error bars. Dashed vertical lines indicate the best-fit power-law index in the case where there is no exponential cutoff, fitting only to FDS3.}
\label{smallbincont}
\end{center}
\end{figure*}

\subsection{Detailed latitude variation of the electron spectrum inferred from gamma rays}
\label{sec:gammalatvariation}

In the gamma rays, unlike the microwaves, we cannot generally neglect the effect of a high-energy break or cutoff in the electron spectrum, as such a break does seem to be preferred by the data \citep{apj2014_14077905}. However, as discussed above, for FDS3 the fit becomes poorly behaved at energies above $\sim 50$ GeV. By keeping only energy bins below $\sim 50$ GeV, we lose information on the high-energy cutoff; accordingly, we will first perform the fit with an unbroken power-law electron spectrum (in practice, we simply set the cutoff to a very high value). Even in this case, however, we still need to compute the full ICS spectrum; in the low-energy Thomson limit, a power-law electron spectrum can be directly translated into a power-law photon spectrum, but this does not hold true when the ICS enters the Klein-Nishina regime.

In this case (electron spectrum modelled as an unbroken power law), we find the following constraints on the power-law index:
\begin{itemize}
\item $-15^\circ < b < -10^\circ$: $p = 2.52 \pm 0.24$
\item $-20^\circ < b < -15^\circ$: $p = 2.70 \pm 0.12$
\item $-25^\circ < b < -20^\circ$: $p = 2.79 \pm 0.09$
\item $-30^\circ < b < -25^\circ$: $p = 2.79 \pm 0.06$
\item $-35^\circ < b < -30^\circ$: $p = 2.70^{+0.03}_{-0.06}$
\end{itemize}

In the data below 50 GeV, there thus appears to be no evidence for a change in the electron spectral index from the gamma-ray data. (In Appendix \ref{app:isrf}, we discuss how this is consistent with the substantial changes in the ISRF with latitude.) The spectra we find in this analysis are systematically slightly softer than those obtained for the FDS1 dataset, where our model was a power-law with a cutoff; this is due to the strong degeneracy between power-law index and cutoff energy, with higher cutoff energies leading to softer power laws (visible in e.g. Fig.~\ref{indcont}). This degeneracy in turn can be easily understood; a lower cutoff energy means fewer high-energy electrons, and thus less high-energy gamma-ray emission relative to lower energies, which can be partly compensated by a harder electron spectrum at energies below the cutoff.

In Fig.~\ref{fig:latcompare} we show the power-law slope inferred by fitting a simple unbroken power law to the microwave data and to FDS3. In general, the two appear broadly consistent at lower latitudes, but FDS3 does not exhibit the preference for a harder power law at high latitudes, especially at $-35^\circ < b < -30^\circ$. This may indicate different origins for the gamma-ray and microwave data, or a hardening spectrum at high latitudes whose effect is masked in the gamma-ray data by the presence of a high-energy cutoff that moves to progressively lower energies at high latitudes. 

To illustrate this latter possibility, in Fig.~\ref{smallbincont} we show the $\Delta \chi^2$ contours in power law index and energy cutoff for each of the FDS3 latitude bands (now modeling the electron spectrum as a power law with an exponential cutoff). We overlay the low-energy power laws preferred by the microwave data in each band (as per Fig.~\ref{fig:latsynch}), and the power law preferred by FDS3 in the absence of a cutoff, to demonstrate how the microwave data may be reconciled with the gamma-ray data by the presence of a cutoff. We see that in the FDS3 dataset (which only extends up to 50 GeV) there are generally two minima, with the second corresponding to a hard power law with relatively low cutoff energy; this second minimum is disfavored when the higher-energy gamma-ray data are included (as in FDS1-2). We see that latitude-by-latitude, the $95\%$ confidence contours for the spectral parameters derived from the gamma-rays overlap with the favored regions from the microwaves, even where the preferred slope for an \emph{unbroken} power law is quite different between the two datasets.

\section{Simultaneous Fits to Microwave and Gamma-Ray Data}
\label{sec:simultaneous}

Since it still appears plausible that a single electron spectrum could explain both datasets (albeit with a harder power law and lower cutoff energy as one moves away from the Galactic plane), we now perform latitude-dependent, simultaneous fits to the microwave and gamma-ray data. We also examine the $\chi^2$/dof as a measure of goodness of fit. Here we use FDS1 and FDS2 in preference to FDS3, as the wider latitude bands reduce the uncertainty (and even so it remains quite substantial). The number of degrees of freedom is the total number of data points (gamma-ray and microwave data combined) minus the number of fit parameters, which unless stated otherwise is 4: $(B, p,E_\mathrm{cut}, A)$, where $A$ controls the overall amplitude.

The results we obtain, using FDS1 and assuming the electron spectrum is a power law with an exponential cutoff, are tabulated above the double line in Table \ref{Results}. We see that the preferred magnetic field value is a few $\mu$G at low latitudes, increases to $\sim 8 \mu$G as $|b|$ increases, and then drops to 0 in the highest latitude bin. The power law value remains approximately $dN/dE \propto E^{-2}$, but prefers a lower value in the mid-latitudes, with a correspondingly lower cutoff energy; this appears to be driven by the hardening of the microwave spectrum in the mid-latitudes, which breaks the degeneracy between power-law slope and cutoff energy in the gamma-ray data.

We can compare these results to those obtained using FDS2 as shown below the double line in Table \ref{Results}. As for FDS1, the magnetic field is between $5-15 \mu$G for $20^\circ < |b| < 40^\circ$ and drops toward zero at higher latitudes. For $|b| > 20^\circ$, the power-law slope is fairly consistent, and similar to the power-law found for the $20-40^\circ$ latitude range with FDS1. At the highest latitudes (where the microwave data can no longer break the degeneracy between cutoff and power-law slope), FDS2 prefers a somewhat lower power law and cutoff in comparison to FDS1, consistent with what we found when verifying previous results for the Bubbles as a whole. However, the results remain broadly consistent.

As discussed previously, the $10-20^\circ$ range in the FDS2 dataset gives peculiar results: the preferred magnetic field (around $4 \mu$G) is consistent with the results of FDS1, but the fit prefers a very steep power-law and low cutoff energy, and the overall fit quality is rather bad. We reiterate that this behavior is likely not physical, but rather due to mismodeling of the background in this band; we discuss this point in detail in Appendix \ref{app:innergalaxy}.

\begin{figure*}
\begin{center}
\begin{tabular}{ll}
\includegraphics[height=0.4\textwidth]{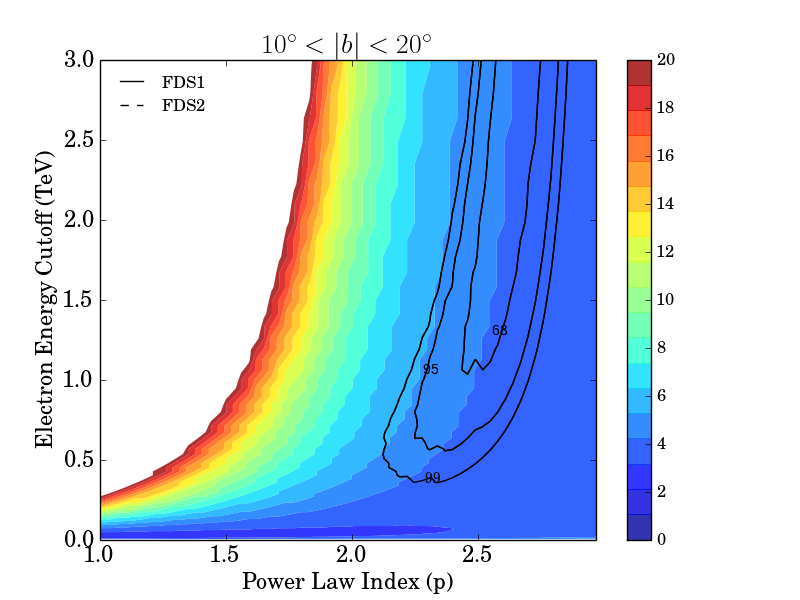} & \includegraphics[height=0.4\textwidth]{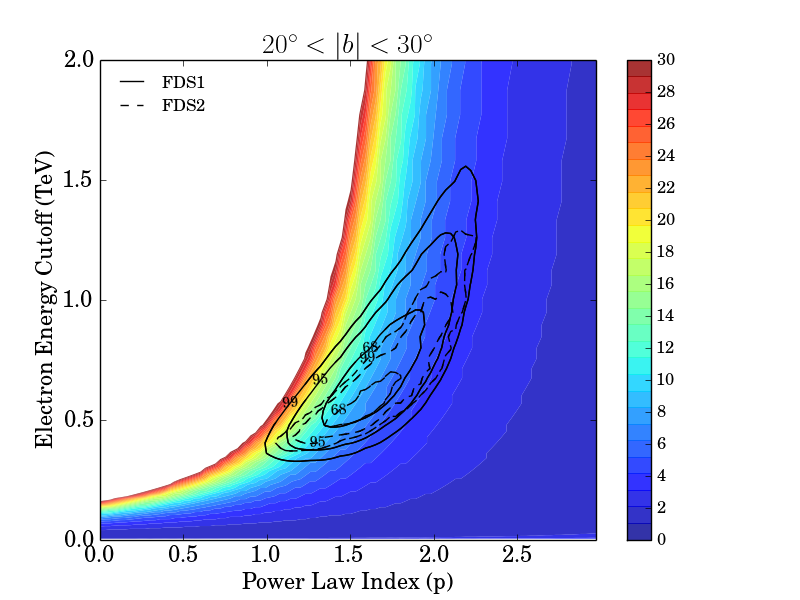} \\ 
\includegraphics[height=0.4\textwidth]{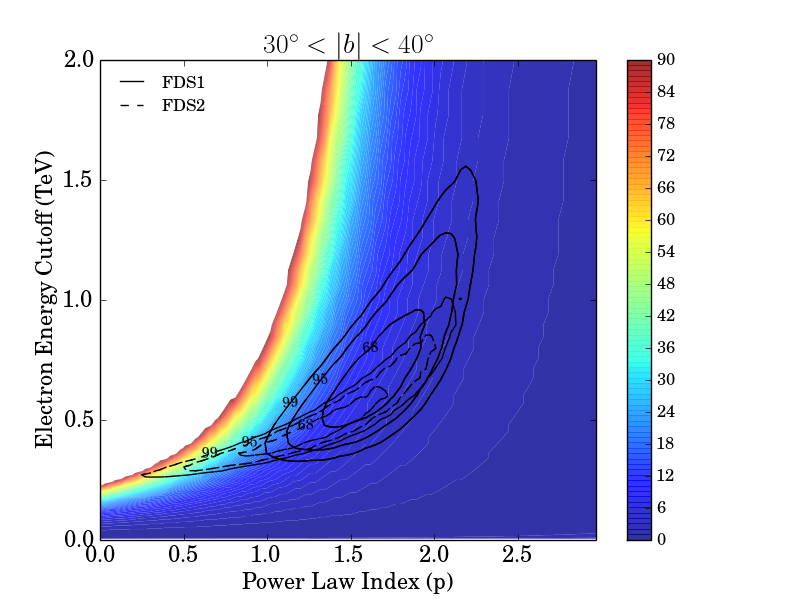} & \includegraphics[height=0.4\textwidth]{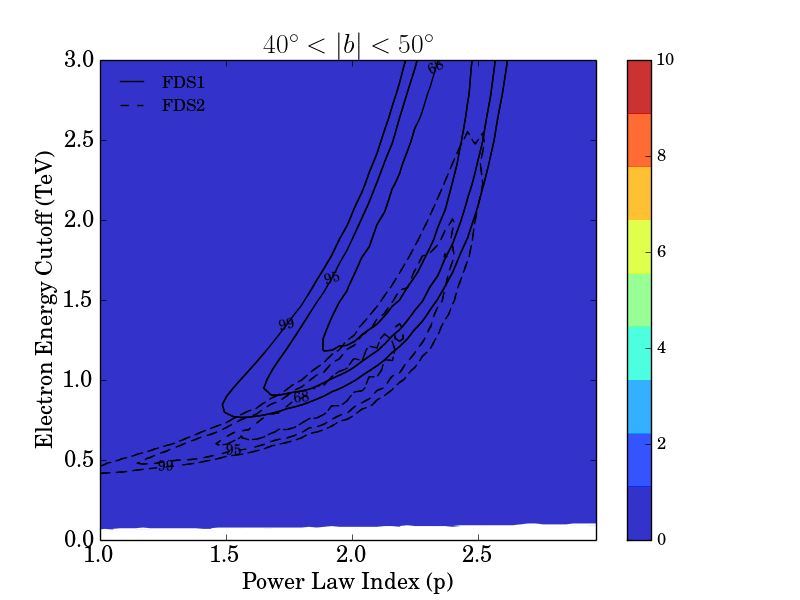} \\ 
\end{tabular}
\caption{Solid and dashed contours describe the $\Delta \chi^2$, relative to the global minimum, for fits to FDS1 (+microwave data) and FDS2 (+microwave data) respectively. Each set of contours has three curves corresponding to confidence intervals of $68$, $95$ and $99$ percent. The coloured contours represent the best-fit magnetic field strengths obtained for each choice of $E_\mathrm{cut}$ and $p$ in the fit including FDS1. In the $20-30^\circ$ and $30-40^\circ$ latitude bands, the FDS1 data for the $20-40^\circ$ interval are used, but the FDS2 data are broken down into the two 10-degree latitude bands for comparison. The electron spectrum model is a power law with an exponential cutoff.
}
\label{overlay_g}
\end{center}
\end{figure*}

\begin{table}
  \setlength\extrarowheight{6pt}
  \begin{center}
  \begin{tabular}{|l|l|l|l|l|}
    \hline
    Latitude Band & B ($\mu$G) & $p$ & $E_\mathrm{cut}$ (TeV) & $\chi^2$/dof \\ \hline
    10-20$^\circ$ & $4.17^{+0.36}_{-0.20}$ & $2.67^{+0.08}_{-0.15}$ & $3.13^{+4.26}_{-1.29}$ & $0.446$ \\ \hline
    20-40$^\circ$ & $7.76^{+2.96}_{-1.71}$ & $1.65^{+0.17}_{-0.21}$ & $0.675^{+0.177}_{-0.126}$ & $0.390$ \\ \hline
    40-60$^\circ$ & $<0.21$ & $2.19^{+0.18}_{-0.18}$ & $1.88^{+0.74}_{-0.44}$ & $1.52$ \\ \hline
    \hline
      10-20$^\circ$ & $4.17^{+0.36}_{-0.20}$ & $<0.23$ & $0.013^{+0.001}_{-0.002}$ & $3.21$ \\ \hline
    20-30$^\circ$ & $13.49^{+3.41}_{-3.10}$ & $1.53^{+0.16}_{-0.12}$ & $0.538^{+0.092}_{-0.032}$ & $1.32$ \\ \hline
    30-40$^\circ$ & $8.91^{+2.31}_{-3.34}$ & $1.20^{+0.39}_{-0.16}$ & $0.428^{+0.135}_{-0.026}$ & $1.20$ \\ \hline
    40-50$^\circ$ & $<0.82$ & $1.86^{+0.21}_{-0.22}$ & $0.847^{+0.272}_{-0.130}$ & $0.787$ \\ \hline
  \end{tabular}
  \caption{Best fit parameters found when fitting a power-law electron spectrum with an exponential cutoff to FDS1 (above the double line) or FDS2 (below the double line), and microwave data from the first three frequency bands of WMAP. Error bars correspond to 68 percent confidence intervals. Upper limits correspond to 95 percent confidence intervals. For FDS1 there are 24 degrees of freedom and for FDS2 there are 29 degrees of freedom.}
    \label{Results}
  \end{center}
\end{table}

Figs. \ref{overlay_g} shows the contours of the \emph{combined} $\Delta \chi^2$ for the fit to FDS1 + microwave data and FDS2 + microwave data, as a function of the power-law index and cutoff energy, marginalizing over the amplitude and the magnetic field strength.  This figure also displays the value of the best-fit magnetic field for each choice of $p$ and $E_\mathrm{cut}$, based on the fit to FDS1 + microwaves, to demonstrate how this varies as a function of power-law and cutoff; the qualitative behavior for the magnetic field determined from the fit to FDS2 is very similar. In general we find good agreement between the results for FDS1 and FDS2 for $|b| > 20^\circ$, as expected from our earlier studies. Even including both microwave and gamma-ray data, there is a residual degeneracy between $p$ and $E_\mathrm{cut}$.

\begin{figure*}
\includegraphics[height=0.35\textwidth]{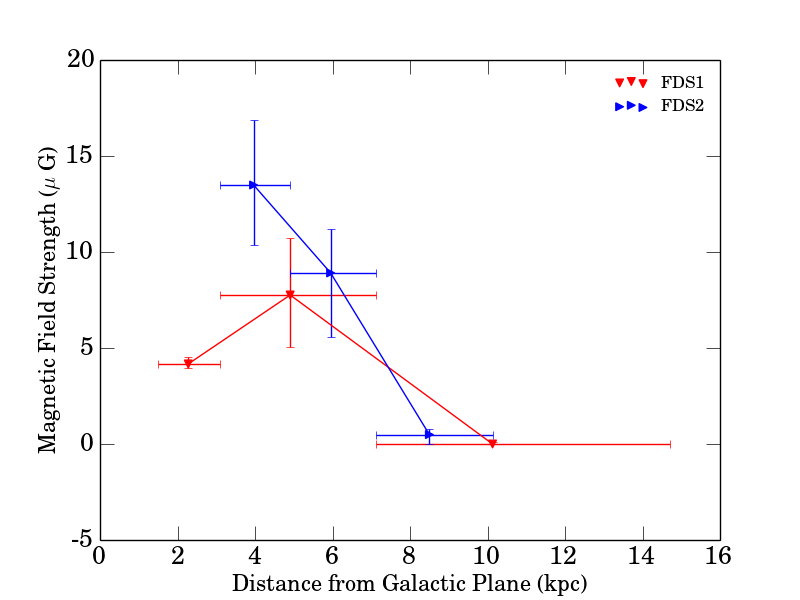}
\includegraphics[height=0.35\textwidth]{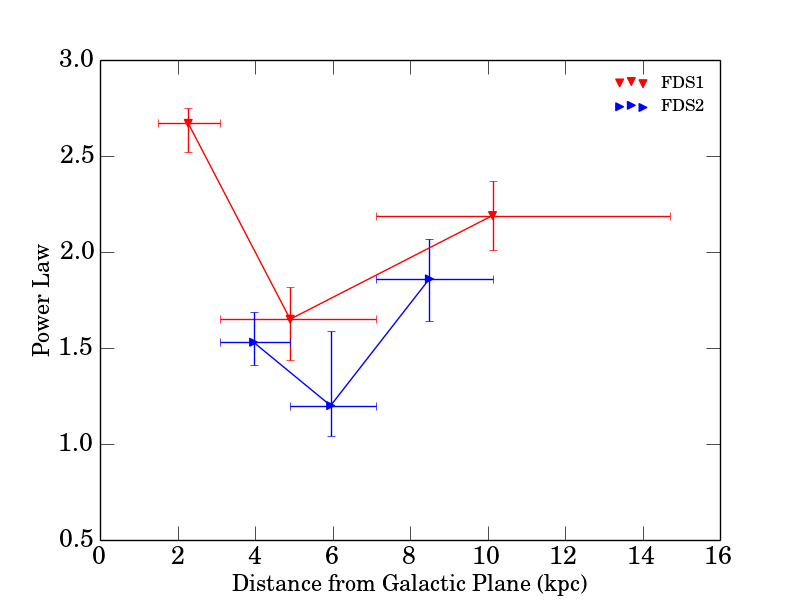}
\caption{The best fit magnetic field strength (left panel) and power law index $p$ (right panel) as a function of distance from the Galactic plane, for two gamma-ray datasets (FDS1 and FDS2), assuming an electron spectrum described by a power law with an exponential cutoff. We do not include results for FDS2 in the lowest latitude bin (see discussion in text).}
\label{allcompb}
\end{figure*}

In Fig.~\ref{allcompb} we summarise these results, plotting the magnetic field and power law strength as a function of height above or below the Galactic plane, where we have taken this quantity to be given by $d_E \tan(b)$, and the distance between the Earth and the GC to be $d_E = 8.5$ kpc. There is a consistent preference for a magnetic field in the range of a few $\mu$G to 15 $\mu$G inside the Bubbles up to $5-7$ kpc from the Galactic plane, falling off at greater distances. There is also some mild indication for an increase in the magnetic field at intermediate latitudes. However, there are clearly systematic uncertainties arising from the gamma-ray dataset used, and (as we will discuss in Appendix \ref{app:hardcutoff}) the details of the high-energy cutoff. In these plots we do not include the results for FDS2 in the $10-20^\circ$ band, since as discussed above (and in detail in Appendix \ref{app:innergalaxy}), the FDS2 data in this latitude band are subject to large systematic uncertainties. If the `inner galaxy' region of interest discussed in Appendix \ref{app:innergalaxy} is used to derive the magnetic field and power law slope for this latitude band, we find results consistent with FDS1.

\begin{figure*}
\begin{center}
\includegraphics[height=0.35\textwidth]{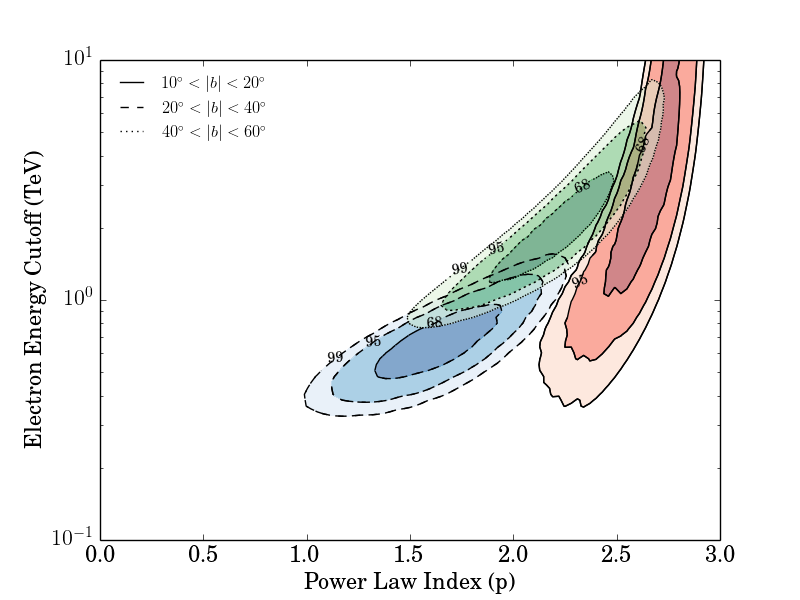}
\includegraphics[height=0.35\textwidth]{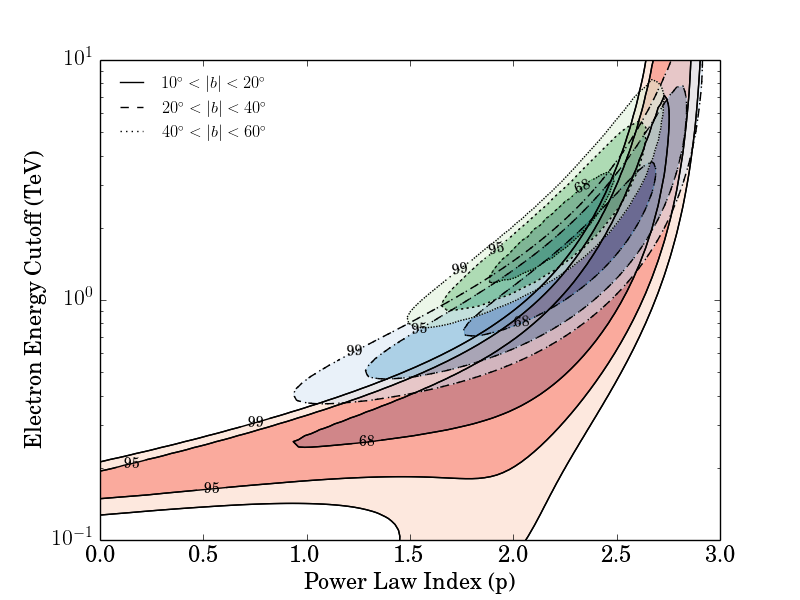}
\caption{We plot the set of three contours, $(68,95,99)$ percent confidence, for each of the three latitude bands in FDS1. The left panel shows the results where the microwave data are included in the fit; the right panel the case where they are not. If we demand consistency with the microwave data under the leptonic-origin hypothesis, the intermediate-latitude region appears to prefer a harder spectrum and lower high-energy cutoff. When the gamma-ray data are fitted independently of the microwaves, the only trend is for a slightly increasing cutoff energy at higher latitudes.}
\label{allcont}
\end{center}
\end{figure*}

The preferred power-law slope shows some evidence for a hardening at intermediate latitudes. In Fig.~\ref{allcont}, we plot the combined $\Delta \chi^2$ contours based on FDS1 and the microwave data, overlaying the results for the three latitude bands of FDS1, to demonstrate this hardening. We also show the fit to FDS1 alone, concatenating results from the panels of Fig.~\ref{indcont}; as discussed above, the microwave data appears to drive the mid-latitude hardening, and when it is not included the only apparent spectral trend is a preference for modestly higher cutoff energies at higher latitudes.

Due to the degeneracies between the magnetic field, cutoff energy and power-law slope, one might wonder if the evidence for a rise in the magnetic field at intermediate latitudes is being driven entirely by the apparent hardening of the spectrum there. We tested a model where the electron spectral parameters $p$ and $E_\mathrm{cut}$, describing a power-law spectrum with an exponential cutoff, were held fixed at the values found in Section \ref{sec:overallcomp}, for \emph{all} latitude bands. Explicitly, we performed a fit to FDS1 with  $p=2.19$ and $E_\mathrm{cut} = 1.62 \TeV$, and to FDS2 with $p=1.89$ and $E_\mathrm{cut} = 0.810 \TeV$. Our results are presented in Table \ref{fds1_fixed}.
\begin{table}
  \setlength\extrarowheight{6pt}
  \begin{center}
  \begin{tabular}{|l|l|l|l|l|}
    \hline
    Latitude Band & B ($\mu$G)  & $\chi^2$/dof \\ \hline
    10-20 & $6.76^{+0.13}_{-0.45}$  & $1.65$ \\ \hline
    20-40 & $4.79^{+0.11}_{-0.33}$  & $0.849$ \\ \hline
    40-60 & $<0.21$  & $1.54$ \\ \hline
    \hline
    10-20 & $16.60^{+0.82}_{-0.75}$  & $3.97$ \\ \hline
    20-30 & $8.32^{+0.32}_{-0.23}$  & $1.40$ \\ \hline
    30-40 & $3.63^{+0.10}_{-0.14}$  & $1.37$ \\ \hline
    40-50 & $<0.73$  & $0.842$ \\ \hline
  \end{tabular}
  \caption{Best fit parameters found when the power-law index $p$ and cutoff energy $E_\mathrm{cut}$ are held fixed with respect to latitude (see text), in fitting a power law electron spectrum with exponential cutoff to FDS1 (above the double lines) or FDS2 (below the double lines), as well as the first three frequencies of the microwave data. Error bars indicate $68$ percent confidence intervals.  For FDS1 there are 24 degrees of freedom and for FDS2 there are 29 degrees of freedom.}
  \label{fds1_fixed}
  \end{center}
\end{table}

In this case we find there is no longer evidence for a rise in the magnetic field at intermediate latitudes; the field strength appears to fall fairly smoothly from $\sim 15 \mu$G to zero as one moves away from the Galactic plane. Furthermore, the $\chi^2$/dof associated with such a fixed-spectrum model is quite good (except in the $10-20^\circ$ range with FDS2, where as discussed above, the data should be treated with caution). Accordingly, while a magnetic field in the few-$10 \mu$G range seems to be consistently required within the Bubbles, up to $5-6$ kpc from the Galactic plane, attempts to map out its variation in detail will need to investigate the systematics associated with the modeling of the electron spectrum.

\section{Conclusions}
\label{sec:conclusion}

We have examined the consistency of the leptonic-origin hypothesis for the \emph{Fermi} Bubbles by studying the electron spectra independently required to reproduce the gamma-ray and microwave spectra of the Bubbles, as a function of latitude. We have reproduced previous results that show broad consistency between the two phenomena. On a latitude-by-latitude basis, there is some evidence for inconsistent behavior -- in particular, the microwave spectrum (and the inferred electron spectrum) appears to harden with increasing distance from the Galactic plane, while the electron spectrum inferred from the gamma-rays shows no such behavior and is consistent with remaining approximately constant in shape. However, a high-energy cutoff in the electron spectrum is preferred by the data and affects the gamma-ray spectrum far more than the microwave one, so the spectra cannot be demonstrated to be significantly discrepant; a consistent solution exists where the spectra grow somewhat harder in the mid-latitudes (explaining the microwave data), while simultaneously having a lower high-energy cutoff (ensuring the gamma-ray spectrum remains nearly unchanged). If this solution cannot be accommodated by models for the Bubbles' origin, that would suggest a hadronic contribution to the gamma-ray spectra (or a latitude-dependent unaccounted-for systematic error in the microwave spectra).

Fitting the microwave and gamma-ray data simultaneously, we find electron spectra and magnetic fields that provide good descriptions of both components, under the hypothesis that they share a purely leptonic origin. The required magnetic fields lie in the range $[0,15] \ \mu$G, while the electron spectrum behaves as a power-law with a high-energy cutoff, and a slope in the range $[1.5,2.5]$ (slope is defined as $p$ where $dN/dE \propto E^{-p}$). The magnetic field appears to drop to essentially zero at latitudes above $40^\circ$, consistent with previous studies. If the apparent hardening of the electron spectrum at intermediate latitudes ($20-40^\circ$) suggested by the microwave data is real, that in turn suggests a rise in the magnetic field in this latitude range; however, if the electron spectrum is constant throughout the Bubbles, we can obtain a fairly good fit to the data with a smoothly falling magnetic field (with increasing latitude).

\vskip 5mm

{\bf Acknowledgements:} This work was supported by the U.S. Department of Energy under grant Contract Numbers DE$-$SC00012567 and DE$-$SC0013999. The authors thank Roland Crocker, Dmitry Malyshev, Lina Necib and Nicholas Rodd for very helpful discussions and comments, and the \emph{Fermi} Collaboration for use of public data. This research made use of the NASA Astrophysics Data System (ADS) and the IDL Astronomy User's Library at Goddard.

\pagebreak

\bibliographystyle{mn2e}
\bibliography{bubbles}

\appendix

\section{A More Detailed Investigation of a Hard-Cutoff Electron Spectrum}
\label{app:hardcutoff}

\begin{figure*}
\begin{center}
\begin{tabular}{ll}
\includegraphics[height=0.35\textwidth]{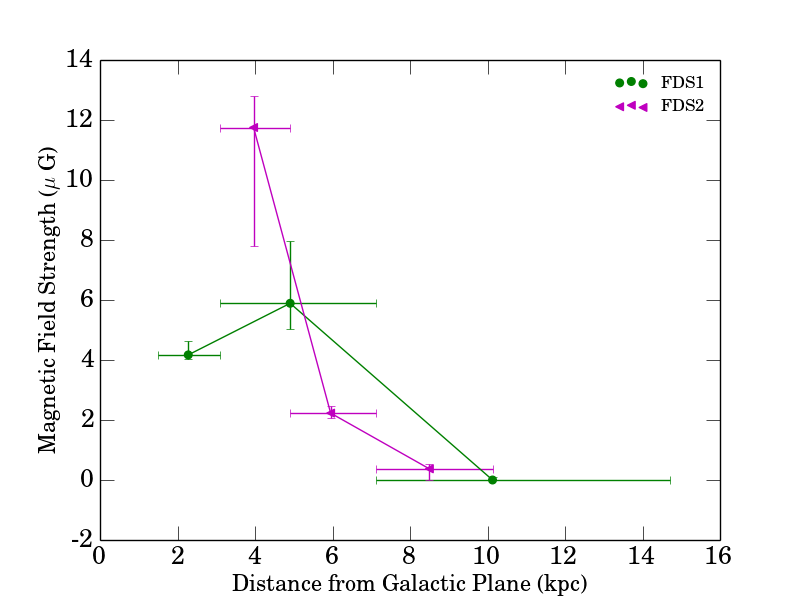} & 
\includegraphics[height=0.35\textwidth]{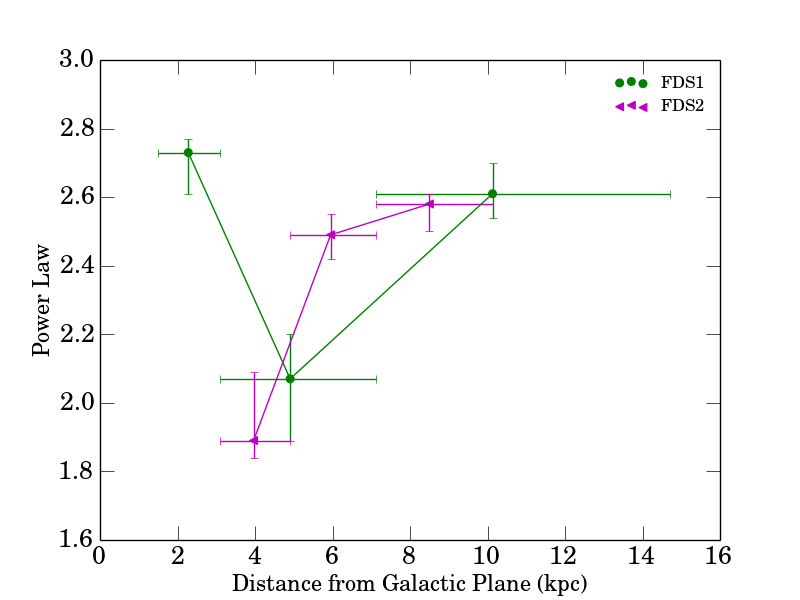}
\end{tabular}
\caption{The best fit magnetic field strength (left panel) and power law index $p$ (right panel) as a function of distance from the Galactic plane, for two gamma-ray datasets (FDS1 and FDS2), assuming an electron spectrum described by a power law with a hard cutoff.}
\label{allcompb2}
\end{center}
\end{figure*}

As discussed in the main text, some earlier studies of the Bubbles have considered models for the electron spectrum with a sharp cutoff above some energy $E_\mathrm{cut}$. Studying such models provides some insight into the sensitivity of our results to the modeling of the electron spectrum around the cutoff point; accordingly, in this appendix we present results for such a model.

Fitting to FDS1 alone in successive latitude bands (the independent fits to the microwave data presumed no cutoff and so do not depend on its modeling), in the $10-20^\circ$ latitude band we find a best-fit power law of $p=2.64^{+0.09}_{-0.12}$, with a high-energy cutoff of $E_\mathrm{cut} = 2.22^{+1.08}_{-0.64}$ TeV. In the $20-40^\circ$ latitude band we obtain $p=2.73^{+0.06}_{-0.09}$ and $E_\mathrm{cut} = 3.71^{+0.95}_{-0.58}$ TeV. In the highest latitude band ($40-60^\circ$), we find $p=2.61^{+0.09}_{-0.06}$ and $E_\mathrm{cut} = 4.15^{+0.77}_{-0.65}$ TeV. We observe that in general a hard cutoff model pushes the power law higher for the gamma ray fits, in comparison to the exponential cutoff model we studied in Section~\ref{sec:latdep}.

\begin{table}
  \setlength\extrarowheight{6pt}
  \begin{center}
  \begin{tabular}{|l|l|l|l|l|}
    \hline
    Latitude Band & B ($\mu$G) & $p$ & $E_\mathrm{cut}$ (TeV) & $\chi^2$/dof \\ \hline
    10-20 & $4.17^{+0.46}_{-0.15}$ & $2.73^{+0.04}_{-0.12}$ & $2.64^{+1.59}_{-0.39}$ & $0.384$ \\ \hline
    20-40 & $5.89^{+2.08}_{-0.85}$ & $2.07^{+0.13}_{-0.18}$ & $1.49^{+0.27}_{-0.24}$ & $1.08$ \\ \hline
    40-60 & $<0.22$ & $2.61^{+0.09}_{-0.07}$ & $4.15^{+0.82}_{-0.61}$ & $1.62$ \\ \hline
    \hline
     10-20 & $14.45^{+1.14}_{-1.18}$ & $2.19^{+0.07}_{-0.03}$ & $1.88^{+0.52}_{-0.08}$ & $3.23$ \\ \hline
    20-30 & $11.75^{+1.06}_{-3.94}$ & $1.89^{+0.20}_{-0.05}$ & $1.06^{+0.16}_{-0.06}$ & $1.36$ \\ \hline
    30-40 & $2.23^{+0.24}_{-0.17}$ & $2.49^{+0.06}_{-0.07}$ & $1.77^{+0.22}_{-0.20}$ & $1.71$ \\ \hline
    40-50 & $<0.54$ & $2.58^{+0.03}_{-0.08}$ & $2.22^{+0.24}_{-0.34}$ & $0.954$ \\ \hline
  \end{tabular}
  \caption{Best fit parameters found when fitting a hard cutoff model to FDS1 (above the double line) or FDS2 (below the double line) and the first three frequencies of WMAP. Error bars correspond to $68$ percent confidence intervals. For FDS1 there are 24 degrees of freedom and for FDS2 there are 29 degrees of freedom.}
  \label{fds1_hard}
  \end{center}
\end{table}

Performing combined fits to the microwave and gamma-ray data, we find the results given in Table \ref{fds1_hard}. Comparing these results to those of Table \ref{Results}, we observe that (as was true for the fit to the gamma-ray data alone) the cutoff energy is generically higher if parametrized by a hard cutoff rather than an exponential cutoff, and the electron spectrum below the cutoff is accordingly softer. The overall quality of fit is generally slightly reduced in these models. In the case of FDS2 in the lowest-latitude band, the hard-cutoff model appears to be finding a completely different minimum to the exponential-cutoff model (with the hard-cutoff model being much more consistent with expectations from FDS1); as we will discuss in Appendix \ref{app:innergalaxy}, this is just one indication of the instability of the fit in this region. Aside from this region, the higher cutoffs and softer power laws found with the hard-cutoff models generally lead to slightly lower preferred values for the magnetic field (since softer power laws mean a larger number of electrons at the energies relevant for WMAP, see the discussion in Section \ref{sec:methods}); however, for FDS1 the values are all consistent within the 68\% confidence limits.

In Fig.~\ref{allcompb2} we plot the best fit power law and magnetic field in the same fashion as in Fig.~\ref{allcompb}, but now assuming a hard cutoff for the power-law electron spectrum. As previously, there is a consistent preference for a magnetic field in the range of a few $\mu$G to 10 $\mu$G inside the Bubbles up to $5-7$ kpc from the Galactic plane, falling off at greater distances, and mild evidence for an increase in the magnetic field at intermediate latitudes. 

The preferred power-law slope is quite sensitive to the modeling of the high-energy cutoff, but we see the familiar trend of a hardening at intermediate latitudes.

\section{The Inner Galaxy: Effect of Background Mismodeling for $|b| < 20^\circ$ in FDS2}
\label{app:innergalaxy}

\begin{figure*}
\begin{center}
\begin{tabular}{ll}
\includegraphics[height=0.4\textwidth]{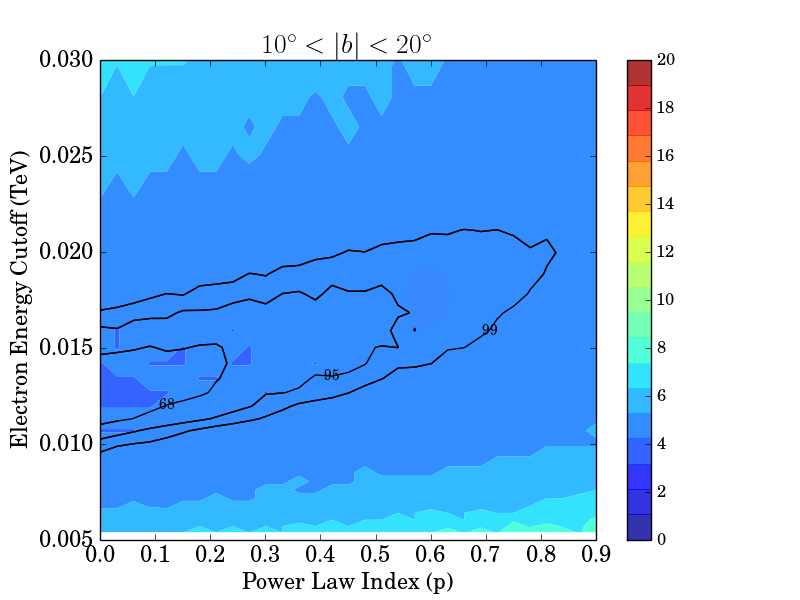} & \includegraphics[height=0.4\textwidth]{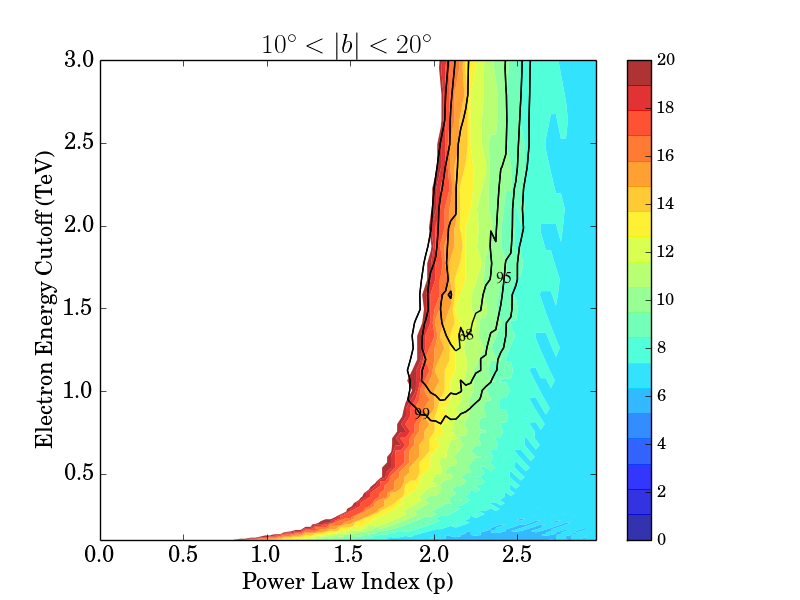} \\
\end{tabular}
\caption{The left panel shows the global minimum for the fit to FDS2 data (overlaid on the best-fit magnetic field contours found using FDS2) in the $10-20^\circ$, from which the results in Table \ref{Results} are derived. The right panel shows the second local minimum, in which the power law index has been forced to be above $0.2$ when determining the minimum $\chi^2$. 
}
\label{1020short}
\end{center}
\end{figure*}

\begin{figure*}
\begin{center}
\begin{tabular}{ll}
\includegraphics[height=0.35\textwidth]{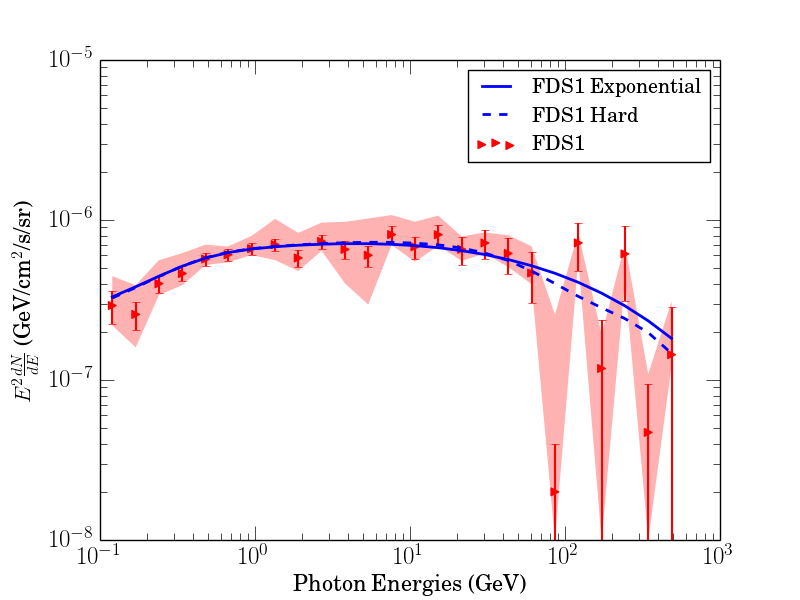} & \includegraphics[height=0.35\textwidth]{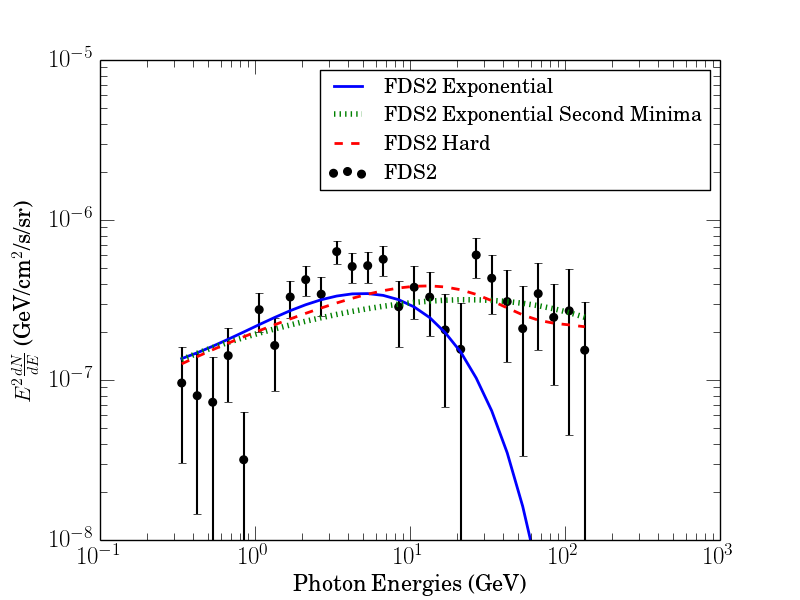} \\
\end{tabular}
\caption{ The left panel shows FDS1 data with the best fit models for a hard cutoff and an exponential cutoff. The right panel shows FDS2 data with the best fit models for a hard cutoff, and both local minima exponential cutoff models.}
\label{1020comp}
\end{center}
\end{figure*}

Comparing the results for the latitude range $|b| = 10-20^\circ$ shown in Table \ref{Results} there appears to be a non-negligible discrepancy between the results obtained from FDS1 and FDS2, with the power law and cutoff preferred by FDS2 seeming much lower than for FDS1. We show regions of Fig.~\ref{overlay_g} in higher magnification in Fig.~\ref{1020short}, for this latitude range and displaying the best-fit magnetic fields derived with FDS2 rather than FDS1, to demonstrate that the $\Delta \chi^2$ surface has two local minima. The global minimum has a very hard power law and very low cutoff energy, while the other local minimum is much closer to the FDS1 results and the results for a power-law spectrum with a sharp cutoff. This behavior does not occur in any of the other latitude ranges we have tested, at least when the gamma-ray data is combined with the synchrotron data, and does not occur in FDS1.

\indent In Fig.~\ref{1020comp} we show FDS1, FDS2 data along with the respective best fit models for an exponential and hard cutoff in the electron spectrum. We see that for FDS1, the exponential and hard cutoff models are very similar. For FDS2 the hard cutoff model and the exponential cutoff model from the second $\chi^2$ minima are essentially the same but the exponential cutoff model corresponding to the global $\chi^2$ minimum deviates significantly from the other curves. In particular, while this model fits the low photon energy data points very well, it then cuts off and fails to match the data at high energies. The higher energy data points have large error bars and so contribute little to the overall $\chi^2$. \\
\indent This low-latitude region is particularly sensitive to mis-subtraction of the diffuse background, due to its proximity to the Galactic plane. The diffuse model is fitted using data from the entire galaxy, but this can result in oversubtraction in the inner galaxy region, as discussed previously in the literature \citep{Daylan:2014rsa, 2015JCAP...03..038C} in the context of the GC excess (in particular, the background modeling by \cite{Daylan:2014rsa} is very similar to that used to obtain FDS2). If we instead study the \emph{Fermi} Bubbles spectrum with the fit only performed in the inner galaxy, $|l|<20^\circ$ (Galactic longitude), $|b|<20^\circ$ (Galactic latitude), then our best fit parameters are as follows
$$B = 3.39^{+0.21}_{-0.03} \mu\mathrm{G}, \ p = 2.82^{+0.02}_{-0.02}, \ E_\mathrm{cut} > 5 \TeV.$$

\begin{figure}
\begin{center}
\includegraphics[height=0.35\textwidth]{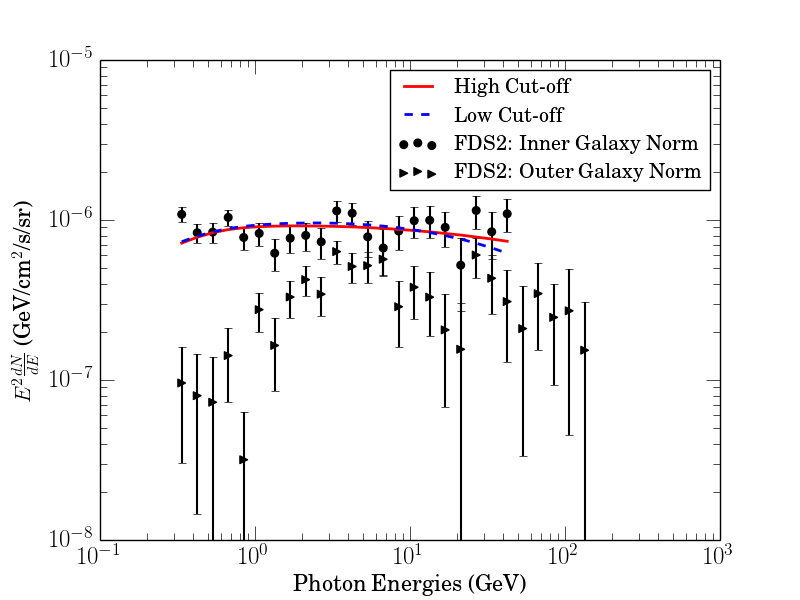}
\caption{We plot FDS2 in the 10-20$^\circ$ latitude range, using both the inner galaxy region of interest (circles) and the full-sky region of interest (triangles). The full-sky result is suppressed at low energies, which we attribute to background oversubtraction.  We also plot the best-fit model to the FDS2 data from the inner galaxy region of interest, as well as the best-fit model when we impose an electron cutoff energy of $1.00 \TeV$.}
\label{1020ingalaxy}
\end{center}
\end{figure}

The best fit values for the magnetic field and the power law now agree more closely with the results obtained for FDS2 with a hard cutoff, as well as to FDS1 for both cutoff models. The cutoff energy, on the other hand, is much higher than previously found. When fitting only within the inner galaxy, the fit becomes unstable at energies above $50 \GeV$; however, omitting data above 50 GeV essentially removes any upper bound on the cutoff energy. If we constrain the cutoff energy to be $1.00 \TeV$, comparable to the cutoff found in the best-fit spectrum to FDS1 in this region, then we obtain a best fit power law of $2.64$ and (as previously) a magnetic field of $3.39 \mu$G. A plot of this model along with the overall best-fit model, with the higher energy cutoff, is shown in Fig.~\ref{1020ingalaxy}. The difference in $\chi^2_\mathrm{ICS}$ between these two models is $3.8$ and the contribution to this difference seems to come mainly from the higher energy data points. The difference in $\chi^2_\mathrm{total}$ is 4.12.

\section{Tests of Some Possible Systematics}
\label{app:systematics}

\begin{figure*}
\begin{center}
\begin{tabular}{ll}
\includegraphics[height=0.4\textwidth]{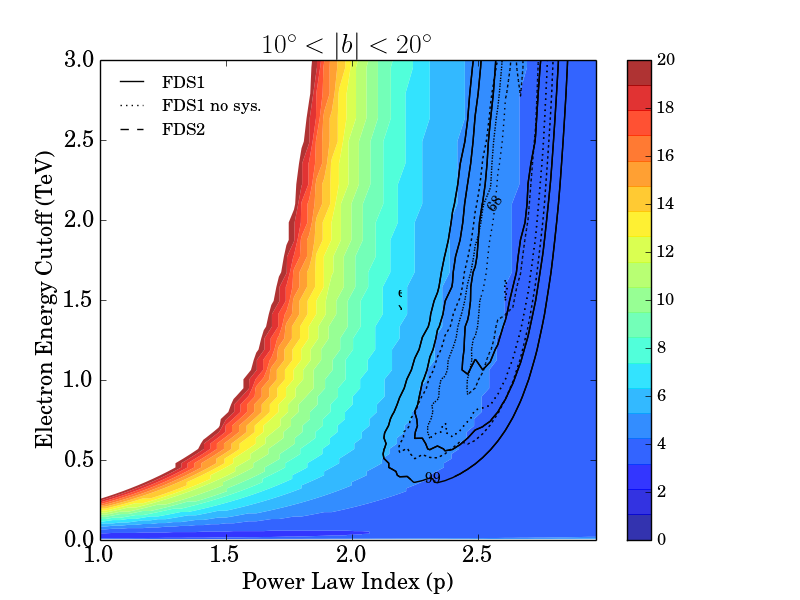} & \includegraphics[height=0.4\textwidth]{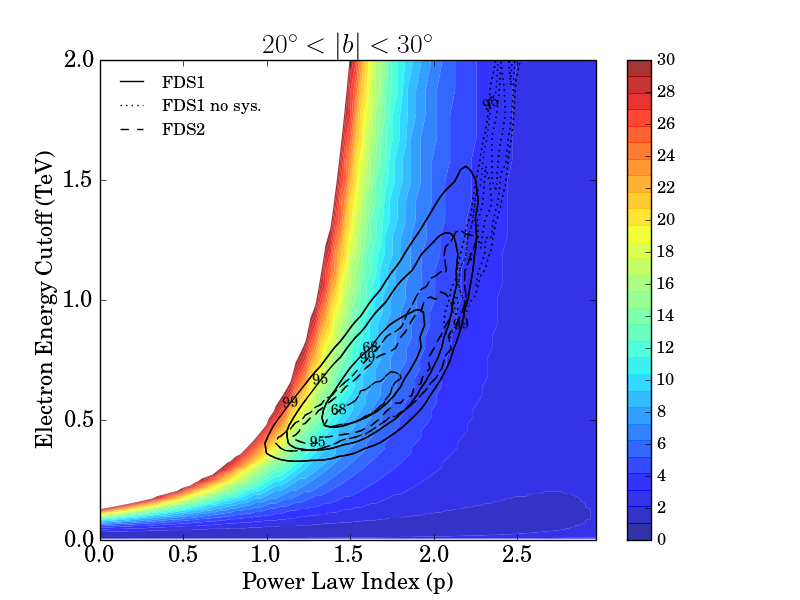} \\ 
\includegraphics[height=0.4\textwidth]{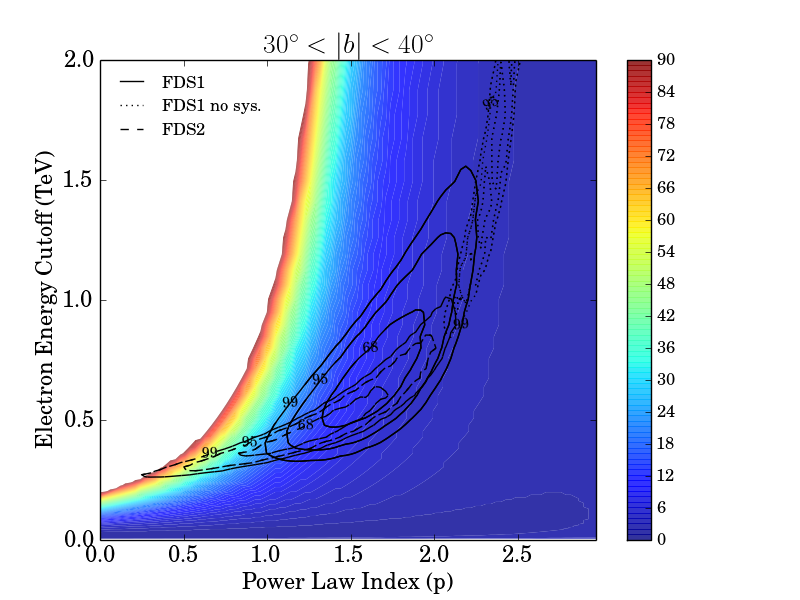} & \includegraphics[height=0.4\textwidth]{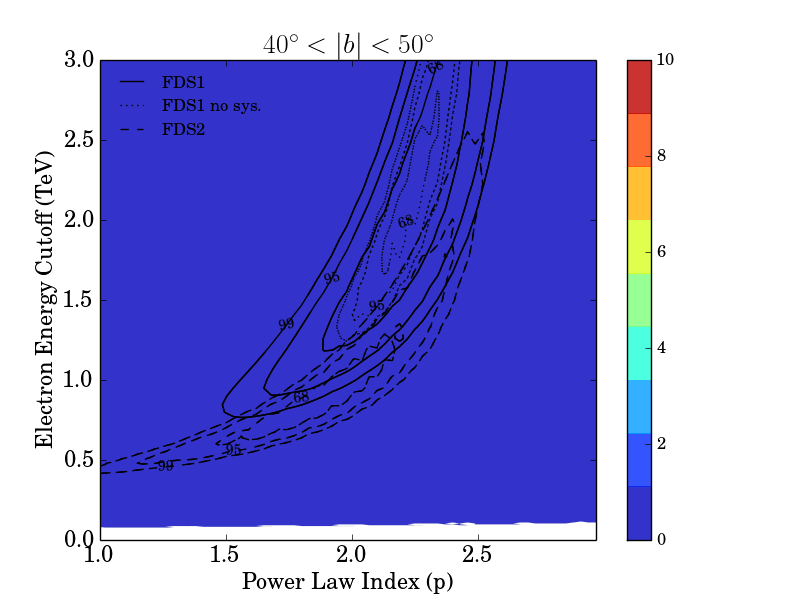} \\ 
\end{tabular}
\caption{As Fig.~\ref{overlay_g}, except that the colour contours now represent the best fit magnetic field strengths obtained in the fit involving FDS1 neglecting systematic uncertainties. The dotted contours are contours of $\Delta \chi^2$ in the same fit.}
\label{overlay_gs}
\end{center}
\end{figure*}

\subsection{Using the Full Microwave Spectrum}

\begin{table}
  \setlength\extrarowheight{6pt}
  \begin{center}
  \begin{tabular}{|l|l|l|l|l|}
    \hline
    Lat. Band & B ($\mu$G) & $p$ & $E_\mathrm{cut}$ (TeV) & $\chi^2$/dof \\ \hline
    10-20 & $3.89^{+0.36}_{-0.11}$ & $2.82^{+0.05}_{-0.15}$ & $9.73^{+15.32}_{-6.57}$ & $1.26$ \\ \hline
    20-40 & $12.59^{+2.97}_{-3.76}$ & $1.35^{+0.19}_{-0.13}$ & $0.538^{+0.131}_{-0.080}$ & $0.664$ \\ \hline
    40-60 & $<0.20$ & $2.19^{+0.18}_{-0.18}$ & $1.88^{+0.74}_{-0.44}$ & $1.80$ \\ \hline
    \hline
    10-20 & $4.79^{+0.39}_{-0.11}$ & $2.82^{+0.05}_{-0.04}$ & $15.33^{+9.73}_{-4.08}$ & $1.67$ \\ \hline
    20-30 & $12.59^{+2.13}_{-2.54}$ & $1.56^{+0.16}_{-0.08}$ & $0.538^{+0.124}_{-0.013}$ & $1.48$ \\ \hline
    30-40 & $14.45^{+15.04}_{-3.23}$ & $0.901^{+0.175}_{-0.367}$ & $0.361^{+0.047}_{-0.065}$ & $1.48$ \\ \hline
    40-50 & $<0.81$ & $1.86^{+0.21}_{-0.22}$ & $0.847^{+0.272}_{-0.130}$ & $0.862$ \\ \hline
  \end{tabular}
  \caption{Best fit parameters found when fitting an exponential cutoff model to FDS1 (above the double line) or FDS2 (below the double line), with all five frequencies of the WMAP Haze spectrum. Error bars indicate $68$ percent confidence intervals.}
  \label{wmap_all_fds1}
  \end{center}
\end{table}

The main analyses in this paper were done using the first three frequencies of the WMAP radiation spectrum. Background noise is larger for the highest two frequencies and results in large systematic error bars that can skew the fit. In this appendix we look at the results obtained if we use all five frequencies as opposed to just the lowest three. Performing the same analyses as in Section \ref{sec:simultaneous}, now with the full five frequencies given by \cite{2012ApJ...750...17D}, we obtain the optimal parameters for each latitude region as given in Table \ref{wmap_all_fds1}.\\

Comparing to Table \ref{Results} we find that adding the last two frequencies does not change the best-fit parameters substantially, although it can slightly reduce the confidence intervals on the parameters. The contribution to the $\chi^2$/dof is also small.

\subsection{Using Only Statistical Errors in FDS1}

Statistical and systematic uncertainties were treated separately in FDS1, but a correlation matrix for the systematic uncertainties was not provided. In our main analysis, we simply added estimates of the systematic and statistical uncertainties in quadrature. Here we consider whether mismodeling of the systematic uncertainties could dramatically change our results; as a simple test, we check the effect of omitting the systematic uncertainties altogether. 

In Table \ref{wmap_fds1_nosys} we present the results obtained using the lowest three frequencies of the microwave spectrum, and using FDS1 with statistical errors only. Comparing these results to Table \ref{Results} we find that the results are comparable within 68 percent confidence. The $\chi^2$/dof has increased, as expected since we have reduced the error bars.

\begin{figure*}
\begin{center}
\begin{tabular}{lll}
\includegraphics[height=0.25\textwidth]{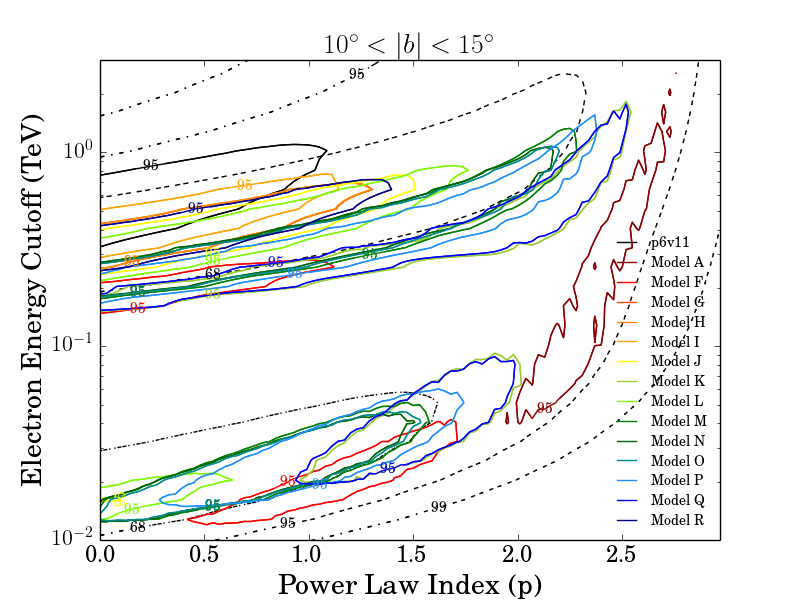} & \includegraphics[height=0.25\textwidth]{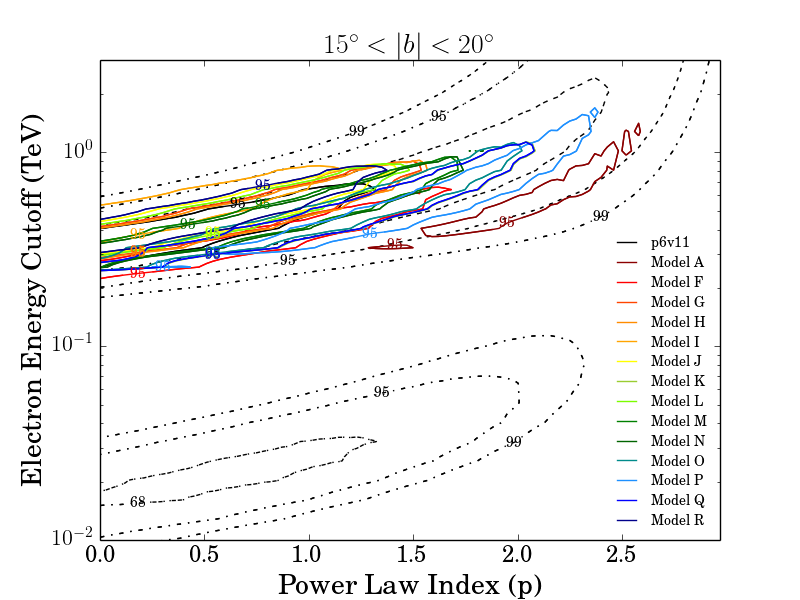} & 
\includegraphics[height=0.25\textwidth]{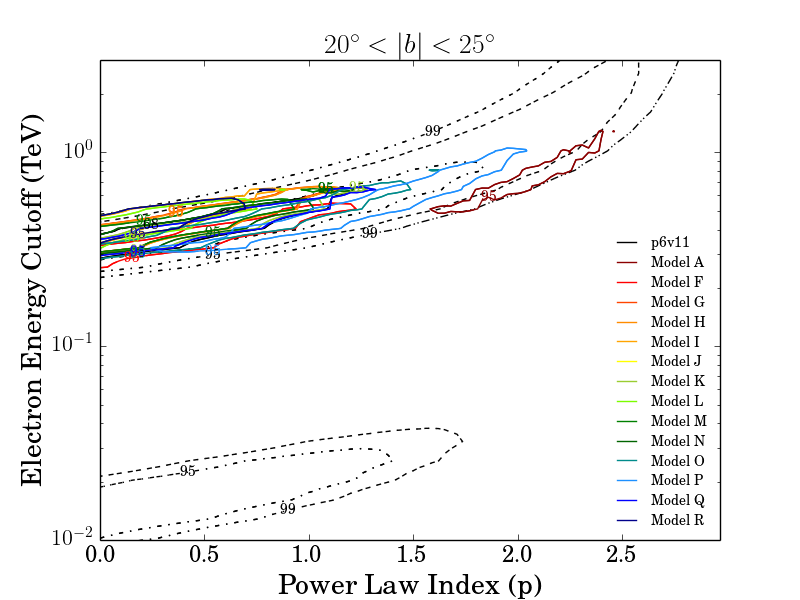}
\end{tabular}
\begin{tabular}{ll}
 \includegraphics[height=0.25\textwidth]{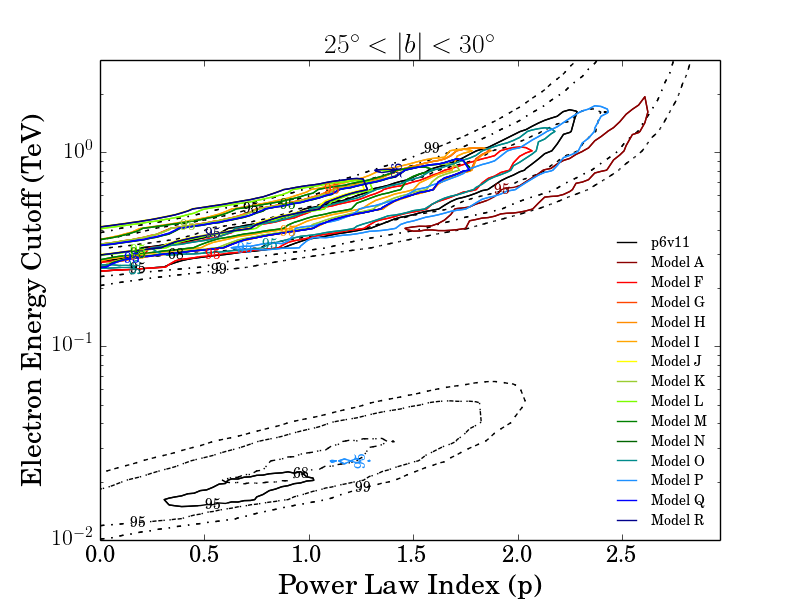} & 
\includegraphics[height=0.25\textwidth]{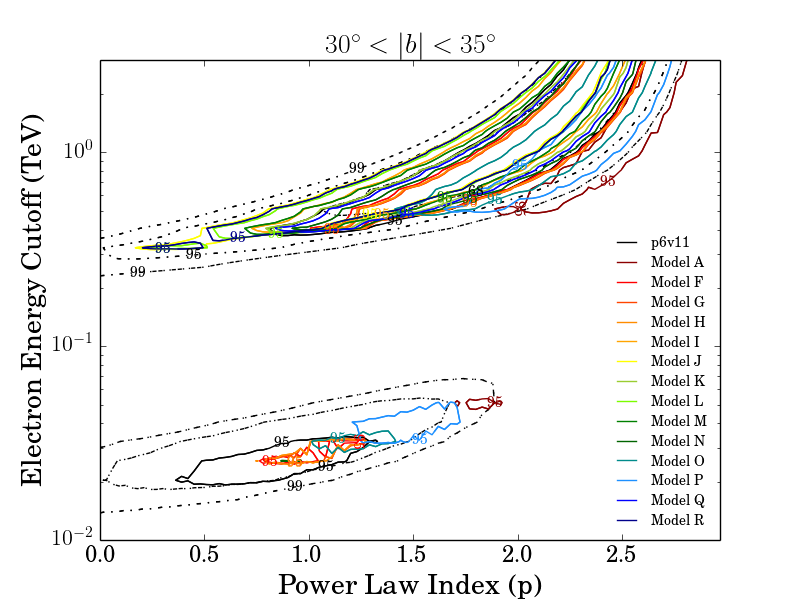} 
\end{tabular}
\caption{We show the favored regions in each of the five latitude bins for FDS3, for a range of different diffuse background models. Contours were obtained from the gamma-ray data only, and are based on a power law electron spectrum with an exponential cutoff. The solid coloured contours are obtained from template fits using 15 different diffuse background models (see text), with statistical errors only; only 95\% confidence contours are shown. The black dashed contours correspond to our standard fit with statistical+systematic errors; 68, 95 and 99\% confidence contours are shown.}
\label{smallbincont2}
\end{center}
\end{figure*}

In Fig.~\ref{overlay_gs}, we show contour plots for $p$ and $E_\mathrm{cut}$ when we include only statistical errors for FDS1, with coloured contours showing the best-fit magnetic field for each choice of $p$, $E_\mathrm{cut}$. In general, the new contours are consistent with (but smaller than) the original FDS1 contours, although at mid-latitudes they prefer slightly softer power laws. This is likely due to the tension, discussed in Section \ref{sec:latdep}, between the softer power laws preferred by gamma-ray data and the harder power laws preferred by the microwave data; reducing the error bars on the gamma-ray data will cause them to exert a larger influence on the fit. The preferred magnetic field strength is consistent in the $10-20^\circ$ and $40-60^\circ$ bands, but is somewhat lower in the $20-40^\circ$ band; this is likely an effect of the softer power law that is preferred when the gamma-ray data is weighted more highly in the fit. However, in both cases the magnetic field remains in the range around $5 \mu\mathrm{G}$ for $|b| \lesssim 40^\circ$.

\begin{table}
  \setlength\extrarowheight{6pt}
  \begin{center}
  \begin{tabular}{|l|l|l|l|l|}
    \hline
    Latitude Band & B ($\mu$G) & $p$ & $E_\mathrm{cut}$ (TeV) & $\chi^2$/dof \\ \hline
    10-20 & $4.47^{+0.16}_{-0.30}$ & $2.58^{+0.13}_{-0.08}$ & $1.77^{+0.74}_{-0.26}$ & $1.52$\\ \hline
    20-40 & $3.39^{+0.32}_{-0.19}$ & $2.37^{+0.03}_{-0.08}$ & $1.58^{+0.35}_{-0.16}$ & $2.23$ \\ \hline
    40-60 & $<0.22$ & $2.25_{-0.02}^{+0.07}$ & $2.22^{+0.19}_{-0.38}$ & $2.33$ \\ \hline
  \end{tabular}
  \caption{Best fit parameters and $68$ percent confidence intervals found when fitting to FDS1 without taking systematic uncertainties into account.}
  \label{wmap_fds1_nosys}
  \end{center}
\end{table}

\subsection{Variation in Diffuse Models in FDS3}

As discussed in the main text, for FDS3 we explore the impact of replacing our default diffuse background model with 14 different \texttt{GALPROP}-based diffuse models. We estimate systematic errors from the scatter in the resulting spectra for the Bubbles. However, one could also perform the full analysis for the different diffuse background models, using statistical errors only, and examine the scatter in the constraints on the electron spectrum as a measure of the systematic error in those constraints. 

In Fig.~\ref{smallbincont2}, we show an example of such an analysis, plotting the $95\%$ confidence level contours for each of the different diffuse models (in colour), as well as the corresponding contours for our standard (full) analysis (black dashed). We see that while the contours for different diffuse models generally overlap, there is considerable variation between the best-fit regions. Our standard analysis generally capture these uncertainties well, with the contours for individual models lying within the corresponding contours for the overall fit. In some cases our standard analysis is perhaps over-conservative, in the sense that it allows regions that are not consistent at high confidence with \emph{any} of the tested diffuse models -- however, the sample we have tested does not represent the full space of possible diffuse background models.

\section{Effects of Scaling the Interstellar Radiation Field}
\label{app:isrf}

\begin{figure*}
\begin{center}
\begin{tabular}{ll}
\includegraphics[height=0.4\textwidth]{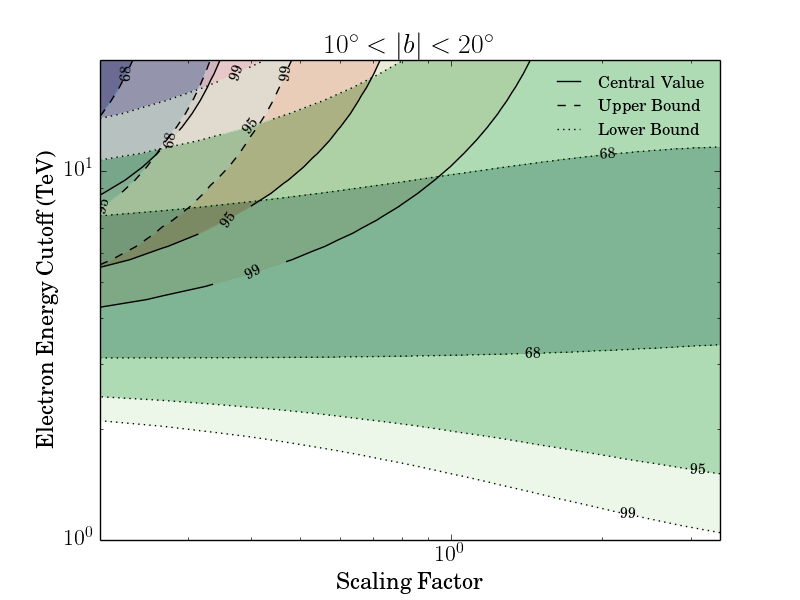} & \includegraphics[height=0.4\textwidth]{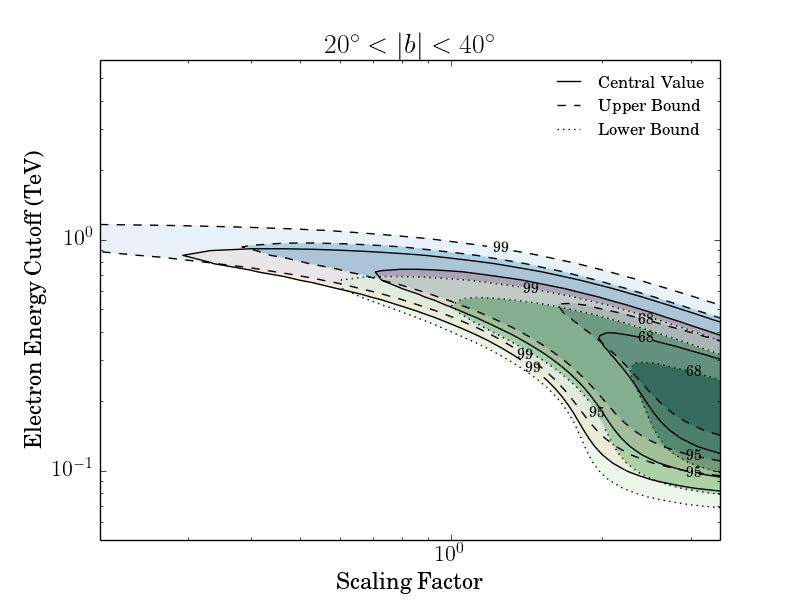} 
\end{tabular}
\caption{We fix $p$ to the value obtained by independently fitting to the WMAP data. Then we marginalize over the electron spectrum normalization, and plot the contours of $\Delta\chi^2$ as a function of ISRF scaling factor (which applies only to the non-CMB part of the ISRF) and $E_\mathrm{cut}$. The green (dotted lines), red (solid lines), and blue (dashed lines) bands are the set of 68, 95, 98 percent confidence contours of $\Delta \chi^2$ when fixing $p$ to its minimum value at 68\% confidence, its central value, and its maximum value at 68\% confidence.
}
\label{floating}
\end{center}
\end{figure*}

\begin{figure*}
\includegraphics[height=0.24\textwidth]{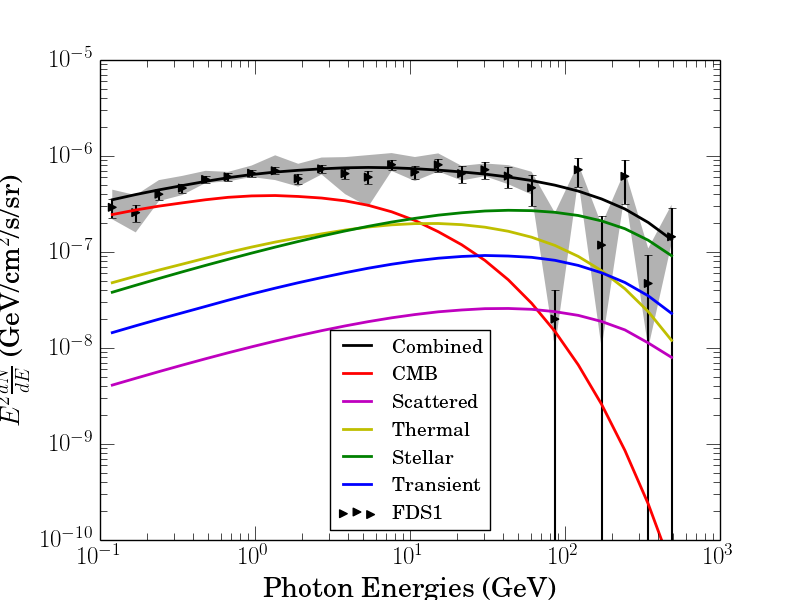} 
\includegraphics[height=0.24\textwidth]{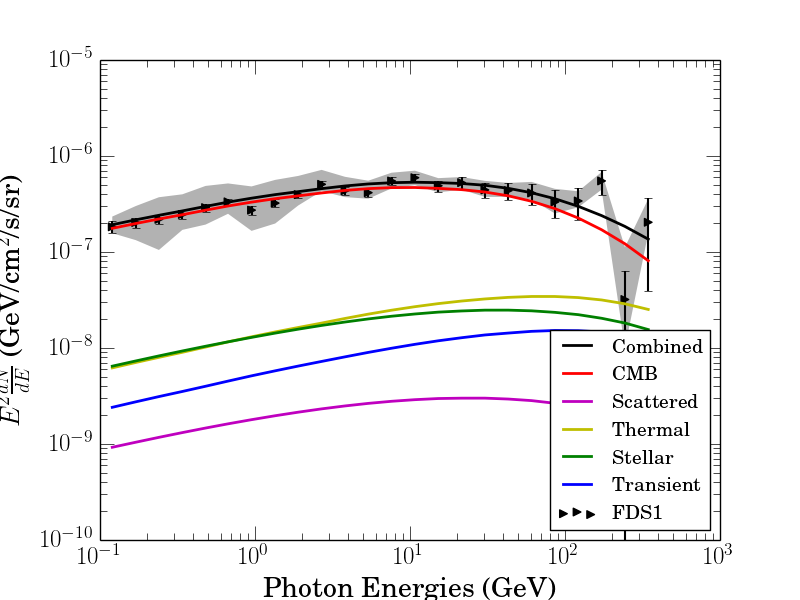}
\includegraphics[height=0.24\textwidth]{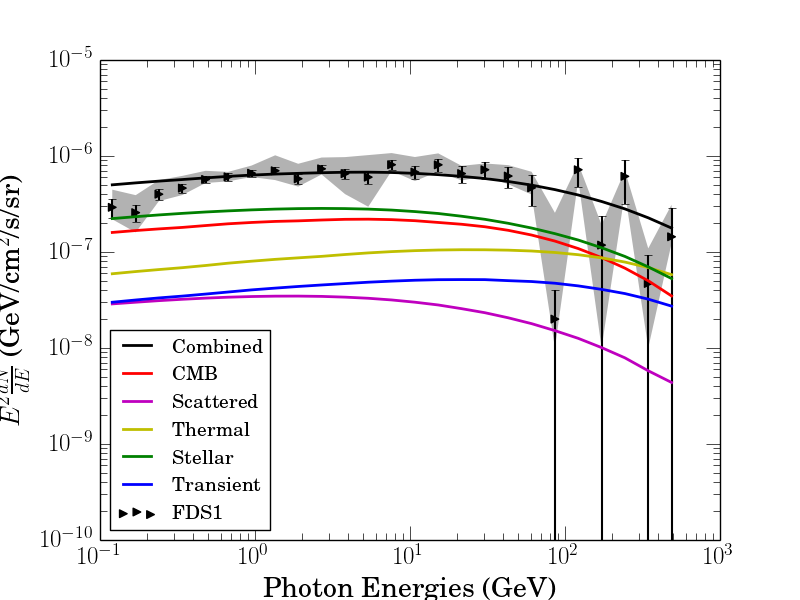}
\caption{Contributions to the overall gamma-ray spectrum from electron scattering on different components of the interstellar radiation field. Black lines show the total spectrum; red, purple, yellow, green, blue lines correspond respectively to scattering on the CMB and the Scattered, Thermal, Stellar and Transient components of the ISRF, as defined in \texttt{GALPROP v54}. The electron spectrum is taken to be the best-fit cutoff power law model from (left panel) FDS1 for $-b=10-20^\circ$; (central panel) FDS1 for $b=-40-60^\circ$; (right panel) FDS1 + synchrotron data for $-b=10-20^\circ$.}
\label{photon_sep}
\end{figure*}

In order to calculate the ICS spectrum we used information about the ISRF and the CMB. One might ask whether allowing the ISRF to vary could affect our results, e.g. by bringing the microwave and gamma-ray spectra into closer agreement. A detailed study of the ISRF uncertainties is beyond the scope of this paper, but we present a simple preliminary analysis here.

In the $10-20^\circ$ and $20-40^\circ$ latitude bins, we took the best-fit power-law obtained from the fit to microwave data (which is unaffected by the ISRF), and performed a fit to the FDS1 data, allowing the normalization of the \emph{non}-CMB contribution to the ISRF to vary as an additional parameter in the fit, along with $E_\mathrm{cut}$ and the electron spectrum amplitude. Explicitly, the power-law values were $p=2.90^{+0.25}_{-0.20}$ and $p=1.55^{+0.15}_{-0.20}$, in the $10-20^\circ$ and the 20-40$^\circ$ latitude bins respectively. We also repeated this test fixing the electron power-law to the values corresponding to the upper and lower $68\%$ confidence bounds derived from the microwave data.

We show the results in Fig.~\ref{floating}. We see that in the $10-20^\circ$ band, if the true power-law were the best-fit value, there would be a preference for an ISRF somewhat fainter than the model we employ (in the context of a leptonic model for the gamma-ray emission); however, a somewhat harder power-law, consistent with the microwave data at $68\%$ confidence, leads to essentially no constraint on the ISRF coefficient, and a consistent preference for an energy cutoff of a few TeV. This is the solution found in our joint fits of Section \ref{sec:simultaneous}.

In the $20-40^\circ$ band, the fit prefers a cutoff around 1 TeV for the standard ISRF normalization and lower ISRF normalizations; high ISRF normalizations would allow a lower energy cutoff, and are slightly preferred, but not at any substantial significance. The preference is stronger for harder power laws. In general we find no strong evidence for a change in (non-CMB) ISRF normalization.

One might also ask, more generally, why the electron spectra extracted from the gamma-ray data for different latitude ranges appear fairly similar (see e.g. Fig.~\ref{allcont}), and the gamma-ray spectra do not seem to exhibit any pronounced variation, even though the ISRF does change markedly in spectrum between latitudes of $10^\circ$ and $50^\circ$. 

As a starting point, consider the case where the spectral model is an unbroken power law, as discussed in Section \ref{sec:gammalatvariation}. In such a case, if we could approximate the ICS process by Thomson scattering, the power-law slope of the resulting gamma-ray emission would be completely independent of the interstellar radiation field. Thus to the degree that the extracted photon spectra are almost invariant with latitude, one would expect the extracted electron spectral indices to also be invariant, consistent with Fig.~\ref{fig:latcompare}.

The approximation of Thomson scattering breaks down when the energy of the upscattered photon becomes comparable to the energy of the incident electron; for the optical part of the ISRF, this occurs for electron energies of $\mathcal{O}(100)$ GeV. Above this energy scale, Klein-Nishina suppression of the scattering cross section causes a softening in the photon spectrum. However, the statistical error bars increase at high energies, so this region usually does not dominate the fit. Furthermore, this effect becomes increasingly unimportant as the energy density in starlight (the highest-frequency part of the ISRF) decreases, i.e. as one moves away from the Galactic plane.

Once the electron spectrum possesses a high-energy cutoff, however, the gamma-ray spectrum is no longer independent of the ISRF spectrum, even in the Thomson regime. Components of the ISRF with different typical energies, such as optical-frequency starlight on one hand and the CMB on the other, give rise to gamma-ray spectra with differing cutoff energies. 

Considering the best-fit models for FDS1 in isolation, in the $10-20^\circ$ and $40-60^\circ$ latitude bands, the gamma-ray spectra are quite similar; the most noticeable difference is that the FDS1 data appear slightly harder at low energies in the high-latitude band, compared to the low-latitude band (see Fig.~\ref{bubbles}). Using the results from Section \ref{sec:latresults}, the best-fit electron spectrum at low latitudes is slightly softer in slope and has a lower energy cutoff, compared to high latitudes. How do these modest-seeming differences in electron spectrum compensate the changes in the ISRF as one moves to higher latitudes?

Examining the contributions from the CMB and the rest of the ISRF to the resulting gamma-ray spectra, we find that in the low-latitude $10-20^\circ$ band, the gamma-ray spectrum is dominated by scattering from the CMB for energies below $\sim 5$ GeV, as shown in the leftmost panel of Figure~\ref{photon_sep}. The relatively low cutoff energy and soft electron spectrum means the CMB contribution drops off sharply above this energy scale. At higher energies, ICS from the starlight in the ISRF dominates the gamma-ray spectrum. The two contributions are of comparable size over a wide range of energies, so the transition is quite smooth.

We plot the corresponding result for high latitudes ($40-60^\circ$) in the central panel of Figure~\ref{photon_sep}; here, in contrast, the CMB contribution dominates everywhere. In order to obtain a spectrum extending to sufficiently high energies, the cutoff energy must be increased relative to low latitudes (the slightly harder electron spectrum below the cutoff also helps in achieving enough high-energy gamma-ray photons, although it is not obvious whether the preference for a harder spectrum comes mainly from this effect or from the low-energy data points).

The best-fit spectrum for low latitudes when the microwave data are included is shown in the rightmost panel of Figure~\ref{photon_sep}, and provides another illustrative example: here the preferred electron spectrum is even softer below the cutoff (in order to match the microwaves), but the cutoff energy is high. In this case the situation is closer to the Thomson limit. The gamma-ray spectra from scattering on the starlight or the CMB are quite similar: the contribution from the CMB is suppressed at very high energies due to the electron energy cutoff, while the contribution from the starlight is suppressed at high energies due to entering the Klein-Nishina regime. The starlight contribution dominates (by an $\mathcal{O}(1)$ factor) at all energies.

Since the gamma-ray spectra produced by the different ISRF components are quite similar in shape -- they share the same low-energy power law, but have varying high-energy cutoffs due to the Klein-Nishina suppression and/or the cutoff in the electron spectrum -- it is straightforward for quite different ISRF models to give rise to very similar gamma-ray spectra, if modest changes to the electron spectral parameters (particularly the cutoff energy) are allowed. The examples above demonstrate that very similar gamma-ray spectra can be obtained in scenarios where scattering on either the CMB or non-CMB ISRF produces most of the visible gamma-rays, as well as the case where both contribute comparably (dominating at different energies).

\end{document}